\documentclass[journal]{IEEEtran}

\ifCLASSINFOpdf
\else
   \usepackage[dvips]{graphicx}
\fi
\usepackage{url}

\usepackage{etoolbox}
\robustify\bfseries
\usepackage{graphicx}
\usepackage{amsmath,amssymb,graphicx}
\usepackage[]{algorithm2e}
\usepackage{tabularx}
\usepackage{wasysym}
\usepackage{color,soul}
\usepackage{cite}
\usepackage{balance}
\usepackage{booktabs}
\usepackage{multirow}
\usepackage{siunitx}
\usepackage{hyperref}
\def\H{{\mathsf H}}
\def\T{{\mathsf T}}
\def\CC{{\mathbb C}}
\def\RR{{\mathbb R}}

\usepackage{makecell}
\usepackage{caption}
\usepackage[font=scriptsize]{subfig}
\newcommand{\subparagraph}{}
\usepackage{titlesec}
\usepackage{textgreek}
\usepackage{enumitem}
\usepackage{arydshln}
\usepackage{fixltx2e}
\usepackage{hyperref}

\usepackage{soul}
\soulregister\cite7 %
\soulregister\citep7 %
\soulregister\citet7 %
\soulregister\ref7 %
\soulregister\pageref7 %
\newcommand{\ZQHL}[1]{#1} %
\newcommand{\ZQHLAfterAccept}[1]{#1} %

\usepackage{flushend} %

\usepackage{amssymb}%
\usepackage{pifont}%

\begin{document}

\title{USDnet: Unsupervised Speech Dereverberation via Neural Forward Filtering}

\author{Zhong-Qiu Wang
\thanks{Z.-Q. Wang is with the Department of Computer Science and Engineering at Southern University of Science and Technology, Shenzhen, Guangdong, China (e-mail: wang.zhongqiu41@gmail.com / wangzq3@sustech.edu.cn).}}

\markboth{}
{Shell \MakeLowercase{\textit{et al.}}: Bare Demo of IEEEtran.cls for IEEE Journals}

\maketitle

\begin{abstract}
In reverberant conditions with a single speaker, each far-field microphone records a reverberant version of the same speaker signal at a different location.
In over-determined conditions, \ZQHL{where there are multiple microphones but only one speaker}, each recorded mixture signal can be leveraged as a constraint to narrow down the solutions to target anechoic speech and thereby reduce reverberation.
Equipped with this insight, we propose USDnet, a novel deep neural network (DNN) approach for unsupervised speech dereverberation (USD).
At each training step, we first feed an input mixture to USDnet to produce an estimate for target speech, and then linearly filter the DNN estimate to approximate the multi-microphone mixture so that the constraint can be satisfied at each microphone, thereby regularizing the DNN estimate to approximate target anechoic speech.
The linear filter can be estimated based on the mixture and DNN estimate via neural forward filtering algorithms such as forward convolutive prediction.
We show that this novel methodology can promote unsupervised dereverberation of single-source reverberant speech.
\end{abstract}


\IEEEpeerreviewmaketitle

\section{Introduction}\label{intro}

\IEEEPARstart{R}{oom} reverberation, caused by signal reflections inside reverberant enclosures, is very detrimental to machine perception in tasks such as automatic speech recognition (ASR) and speaker recognition \cite{A.P.Habets2018}.
It also dramatically degrades human hearing, leading to much lower speech intelligibility and quality.
Rereverberation results from a linear convolution between a room impulse response and the source signal, and speech dereverberation is known as a classic blind deconvolution problem \cite{Levin2011, A.P.Habets2018, BlindDeconvWiki}, which is ill-posed in nature and difficult to be solved as only the convolved signal is observed but the filter and source are both unknown and need to be estimated.
There are infinite solutions to the filter and source pair, where, in each solution, their linear-convolution result can well-approximate the observed reverberant speech.
For example, even if the source estimate is a reverberant version of the source, there still exists a filter such that convolving the source estimate with the filter can roughly approximate the mixture.
The key to successful speech dereverberation, we believe, is to cleverly narrow down the infinite solutions to the source and filter so that the models or algorithms can have a clear idea about what target signals to predict.

A popular approach for speech dereverberation is based on inverse filtering \cite{A.P.Habets2018, Yoshioka2012, Gannot2017, Haeb-Umbach2020}, mainly leveraging signal processing principles.
Representative algorithms include weighted prediction error (WPE) \cite{Yoshioka2009, Yoshioka2010Thesis, Nakatani2010, Jukiac2015, Jukic2017}, where a linear filter is computed to filter past observations with a positive prediction delay to estimate late reverberation to remove.
The rationale is that the speech signals beyond the prediction delay in past frames is little linearly-correlated with the target anechoic signal at the current frame.
WPE leverages this physical principle to narrow down the solutions to target speech in an unsupervised way.
It has gradually become the most popular and successful dereverberation algorithm \cite{Haeb-Umbach2020}, often producing consistent improvement for modern ASR systems.
Subsequent studies along this line of research combine WPE with beamformers in a sequential \cite{Haeb-Umbach2020} or in a jointly-optimal \cite{NakataniCB2019, Boeddeker2020} way.
\ZQHL{It should be noted that the dereveberation results of WPE typically contain some early reflections due to its positive prediction delay.
This is however often acceptable in many applications.
For example, early reflections are found beneficial for speech intelligibility \cite{Moore2012Book}.}

Another popular approach is based on supervised learning, where anechoic and reverberant speech pairs are synthesized via room simulation and a supervised DNN is trained to predict the anechoic speech from reverberant speech \cite{Han2015,  May2017, Ernst2018, Zhao2018c, Zhao2020, Wang2020CSMDereverbJournal, Wang2020d, Kothapally2022, Lemercier2023, Kothapally2024}.
\ZQHL{This approach usually requires a large amount of simulated training data to avoid potential issues related to data scarcity \cite{Alzubaidi2023DataScarcity}.}
The target anechoic speech serves as the training label, and the DNN can learn to find the true solution to target speech through supervised learning.
Recent studies along this line of research leverage target speech estimated by supervised DNNs to improve signal processing based dereverberation \cite{Kinoshita2017, Heymann2018b, Heymann2019, Drude2018a, Zhang2020, Wang2021FCPjournal}, and leverage generative modeling \cite{Richter2023} to improve dereverberation.
This approach is often effective as the DNN has strong capabilities at modeling speech patterns that can be very informative for dereverberation.

In this context, we propose to investigate unsupervised neural speech dereverberation.
Different from signal processing based approaches, which are unsupervised in nature and usually not designed to learn speech patterns from massive training data, we leverage unsupervised deep learning to model speech patterns.
Different from supervised learning based approaches, unsupervised neural dereverberation models can be trained directly on real-recorded reverberant speech, avoiding using simulated training data which could incur generalization issues \cite{Barker2018, Watanabe2020, Zhang2021, Tzinis2022REMIXT, Leglaive2023}.

The key to unsupervised neural speech dereverberation, we believe, is to cleverly narrow down the infinite solutions to the source and filter so that DNNs can receive a clear \textit{supervision} on what target signals to predict, or a clear \textit{regularization} on what estimates produced by the DNN are not good and should be avoided.
This way, the strong modeling capability of DNNs on speech signals can be sufficiently utilized for dereverberation, rather than being wasted due to obscure supervision.
Motivated by the UNSSOR algorithm \cite{Wang2023UNSSOR}, which deals with unsupervised neural speaker separation and does not perform any dereverberation, we assume the availability of over-determined training mixtures (i.e., multi-channel reverberant speech in single-speaker cases), and leverage each mixture signal, during training, as a \textit{mixture constraint} to narrow down the solutions to the source and filter.
This way, the DNN can (a) receive a clear \textit{supervision} about what the target speech is by checking whether its prediction satisfies the mixture constraints during training; and (b) be trained directly on a set of collected reverberant signals to realize dereverberation.
\ZQHL{Our contributions are summarized as follows}:
\begin{itemize}[leftmargin=*,noitemsep,topsep=0pt]
\item We, for the first time, formulate unsupervised neural speech dereverberation as a blind deconvolution problem, which requires estimating both the source and the filter.
\item We propose USDnet, which estimates target (source) speech by modeling speech patterns via unsupervised deep learning.
Based on the source estimate and input mixture, filter estimation is formulated as a linear regression problem, which can be solved via a neural forward filtering algorithm named forward convolutive prediction (FCP) \cite{Wang2021FCPjournal} in multiple ways.
\item Following the blind deconvolution problem, we propose novel mixture-constraint loss functions. \ZQHLAfterAccept{They leverage multi-microphone mixtures as constraints to regularize the DNN estimate.
Together with several other regularizations (which we will introduce), they are found effective at leading the DNN estimate to approximate target speech (and realize dereverberation).}
\end{itemize}
Evaluation results on two speech dereverberation tasks show that USDnet can effectively reduce reverberation.
A sound demo page is provided in the link below.\footnote{See \href{https://zqwang7.github.io/demos/USDnet_demo/index.html}{https://zqwang7.github.io/demos/USDnet\_demo/index.html}.}

\section{Related Work}

Unsupervised neural speech separation, mainly being studied for speech de-noising, separating mixture of multiple anechoic speakers, and separating multi-speaker mixtures to reverberant speaker images without any dereverberation, has attracted consistent research interests in recent years.
However, unsupervised neural dereverberation has been largely under-explored, although dereverberation is an important task in speech separation and enhancement.

MixIT \cite{Wisdom2020MixIT, Tzinis2022REMIXT, Saijo2023ReMixIT}, a representative algorithm, first synthesizes mixtures of mixtures (MoM), each created by mixing existing recorded mixtures, and then trains a DNN to separate each MoM to a number of sources such that the separated sources can be partitioned into groups and, within each group, the separated sources can sum up to one of the mixtures used for synthesizing the MoM. 
Although MixIT is shown capable of separating mixtures of independent reverberant sources, it cannot reduce the reverberation of each source as the reflected signals of each source are dependent with each other.
Noisy-target training \cite{Fujimura2021, Fujimura2023} is an unsupervised approach for speech de-noising.
Similarly to MixIT, it synthetically mixes noisy speech with noise, and trains DNNs to predict the noisy speech based on the synthetic mixtures.
This approach is found capable of reducing noise as speech and noise are independent.
However, whether it would be effective at dereverberation is unknown, especially when dealing with reflection signals that are highly dependent with each other.

Another approach requires source prior models, which can be built based on clean anechoic speech \cite{Nakagome2021, Bie2022, Hao2023, Saito2023DEREVERB}, or anechoic and reverberant speech with some annotations \cite{Fu2022MetricGANU}.
The prior models can, to some extent, narrow down the solutions to target speech and provide a supervision on whether DNN estimates are desired or not during training or inference.

There are approaches leveraging pseudo-labels produced by conventional algorithms such as spatial clustering and independent vector analysis (IVA) for training unsupervised neural source separation models \cite{Tzinis2019, Drude2019UnsupervisedDC, Seetharaman2018UnsupervisedDC, Togami2020UnsupervisedLGM}, leveraging pseudo-labels provided by DNN-supported WPE for training unsupervised dereverberation models \cite{Petkov2019}, or training DNNs to optimize the objective functions of conventional separation algorithms \cite{Drude2019UnsupervisedMaskBeamforming, Bando2021NeuralFCA, Wang2023RVAEEM}.
Their systems heavily rely on conventional algorithms and their performance is often limited by conventional algorithms.

There are studies realizing unsupervised speech de-noising via positive and unlabeled learning \cite{Ito2023PULearning}, where noisy speech and noise signals are assumed available so that a binary classifier can be trained to determine whether each small patch of T-F units belongs to noisy speech that can be maintained or noise that should be removed.
However, it is unclear whether this approach can be modified for dereverberation.

The proposed USDnet is motivated by UNSSOR \cite{Wang2023UNSSOR}, which formulates unsupervised neural speaker separation as a blind deconvolution problem, but UNSSOR is designed to maintain the reverberation of each speaker rather than reducing it.
As we will show in this paper, unsupervised dereverberation has unique challenges and difficulties very different from the ones in unsupervised separation of mixed reverberant speakers, and therefore calls for innovative solutions.

During training, USDnet utilizes recorded mixtures as mixture constraints to narrow down target speech and realizes unsupervised dereverberation by solely leveraging recorded mixtures.
It exploits implicit regularization afforded by multi-microphone signals, and does not require using synthetic data (e.g., MoM, and paired clean and corrupted speech), source prior models, or pseudo-labels provided by conventional algorithms.
These differences distinguish USDnet from existing unsupervised neural methods, and, meanwhile, indicate that USDnet can complement many existing algorithms.

\section{Problem Formulation}\label{problem_formulation}

\ZQHL{This section describes the hypothesized physical models, formulates speech dereverberation as a blind deconvolution problem, highlights the benefits of over-determined conditions, and overviews the proposed USDnet.}

\subsection{Physical Models}

For a $P$-microphone mixture with a single speaker in noisy-reverberant conditions, the physical model can be formulated in the short-time Fourier transform (STFT) domain as follows.

At a reference microphone $q\in \{1,\dots,P\}$, we have
\begin{align} 
	Y_q(t,f) &= X_q(t,f) + \varepsilon_q(t,f) \nonumber \\ 
                 &= S_q(t,f) + R_q(t,f) + \varepsilon_q(t,f) \nonumber \\
                 &= S_q(t,f) + \mathbf{g}_q(f)^\H \widetilde{\mathbf{S}}_q(t,f) + \varepsilon'_q(t,f),
\label{phymodel_freq_reference}
\end{align}
where $t$ indexes $T$ frames and $f$ indexes $F$ frequencies.
In the first row, $Y_q(t,f)$, $X_q(t,f)$, and $\varepsilon_q(t,f)$ respectively denote the STFT coefficients of the noisy-reverberant mixture, reverberant speaker image, and a weak noise signal captured at time $t$, frequency $f$ and microphone $q$.
In the rest of this paper, when dropping indices $t$, $f$ and $q$, we refer to the corresponding spectrograms.
In the second row, we decompose $X_q$ into direct-path signal $S_q$ and reverberation $R_q$ (i.e., $X_q=S_q + R_q$).
In the third row, following narrowband approximation \cite{Talmon2009CTF, Gannot2017} we model reverberation $R_q$ as a linear convolution between a linear filter $\mathbf{g}_q$ and the source $S_q$, i.e., $R_q(t,f)\approx\mathbf{g}_q(f)^\H \widetilde{\mathbf{S}}_q(t,f)$, where $\widetilde{S}_q(t,f) = \big[ S_q(t-K+1,f),\dots,S_q(t-\Delta,f)\big]^\T \in \CC^{K-\Delta}$ stacks a window of T-F units, $\Delta$ is a non-negative prediction delay, $\mathbf{g}_q(f)\in \CC^{K-\Delta}$ is a relative transfer function relating direct-path $S_q(\cdot, f)$ to reverberation $R_q(\cdot, f)$, $(\cdot)^\H$ computes Hermitian transpose, and $\varepsilon'_q$ absorbs the modeling error.

At any non-reference microphone $p\in \{1,\dots,P\}$, where $p\neq q$, we formulate the physical model as
\begin{align} 
	Y_p(t,f) &= X_p(t,f) + \varepsilon_p(t,f) \nonumber \\ 
                 &= \mathbf{h}_p(f)^\H \overline{\mathbf{S}}_q(t,f) + \varepsilon'_p(t,f),
\label{phymodel_freq_non_reference}
\end{align}
where $\overline{\mathbf{S}}_q(t,f) = \big[ S_q(t-I+1,f),\dots,S_q(t+J,f)\big]^\T \in \CC^{I+J}$ stacks a window of T-F units, $\mathbf{h}_p(f) \in \CC^{I+J}$ is the relative transfer function relating $S_q$ at the reference microphone $q$ to the reverberant image at non-reference microphone $p$ (i.e., $X_p$), and $\varepsilon'_p$ absorbs the modeling error.

\subsection{Formulating Dereverberation as Blind Deconvolution}\label{formulation_as_blind_deconv}

We aim at reducing the reverberation of the target speaker in an unsupervised way.
With the physical models in (\ref{phymodel_freq_reference}) and (\ref{phymodel_freq_non_reference}), unsupervised dereverberation can be realized by solving, e.g., the minimization problem below:
\begin{align}
&\underset{\mathbf{g}_{\cdot}(\cdot),\mathbf{h}_{\cdot}(\cdot),S_q(\cdot,\cdot)}{\text{argmin}}
\Big( \sum_{t,f} \Big| Y_q(t,f) - S_q(t,f) - \mathbf{g}_q(f)^\H \widetilde{\mathbf{S}}_q(t,f) \Big|^2 \nonumber \\
&\quad\quad\quad\quad+ \sum_{p=1, p\neq q}^{P} \sum_{t,f} \Big| Y_p(t,f) - \mathbf{h}_p(f)^\H \overline{\mathbf{S}}_q(t,f) \Big|^2\Big), \label{ideal_loss}
\end{align}
which finds source and filters that are most consistent with the physical models in (\ref{phymodel_freq_reference}) and (\ref{phymodel_freq_non_reference}) (i.e., most satisfied with the mixture constraints).
This is a blind deconvolution problem \cite{Levin2011}, which is non-convex in nature and is not solvable if not assuming any prior knowledge about the filters or the source, because all of them are unknown and need to be estimated.

\ZQHL{
\subsection{Benefits of Over-determined Conditions}\label{over-determined_disucussion}

We point out that the more microphones we have, the more likely we could solve the problem in (\ref{ideal_loss}).
To provide an intuitive interpretation, following UNSSOR \cite{Wang2023UNSSOR} we can view the physical models in (\ref{phymodel_freq_reference}) and (\ref{phymodel_freq_non_reference}) as a system of linear equations with many unknowns, where each linear equation can be considered as a constraint to some of the unknowns.
When certain conditions (which we will describe next) are satisfied, the equations can significantly outnumber unknowns, making it more likely to successfully estimate the unknowns.

In detail, suppose that the modeling error $\varepsilon'$ is negligible.
The linear system in (\ref{phymodel_freq_reference}) and (\ref{phymodel_freq_non_reference}) has in total $P\times T\times F$ linear equations, as we have a mixture observation for each $Y_p(t,f)$.
Meanwhile, it has $T\times F + (K-\Delta)\times F + (I+J)\times (P-1) \times F$ unknowns, where the first term, $T\times F$, is because we have an unknown for each $S_q(t,f)$, the second term, $(K-\Delta)\times F$, is because we have $K-\Delta$ unknowns for $\mathbf{g}_q(f)\in \CC^{K-\Delta}$ at the reference microphone $q$ and each frequency $f$, and the third term, $(I+J)\times (P-1) \times F$, is because we have $I+J$ unknowns for each $\mathbf{h}_p(f) \in \CC^{I+J}$ at each non-reference microphone $p$ and each frequency $f$.

In many cases, the linear filters are usually shorter than one second (meaning that $K-\Delta$ and $I+J$ can be assumed small).
If the recording is reasonably long (i.e., $T\gg K-\Delta$ and $T\gg I+J$) and when there are multiple microphones (i.e., $P>1$), the number of equations can be much larger than the number of unknowns.
That is, $P\times T\times F \gg T\times F + (K-\Delta)\times F + (I+J)\times (P-1) \times F$) when $P>1$.
In other words, we consider each mixture observation as a constraint which the estimate of target speech has to satisfy. 
When there are many more constraints than unknowns, we can more likely narrow down the solutions to target speech.

On the other hand, if we only have one microphone (i.e., $P=1$), the number of equations, $T\times F$, is smaller than the number of unknowns, $T\times F + (K-\Delta)\times F$.
In this case, it would be more difficult to pinpoint the true solution to target speech using the approach in (\ref{ideal_loss}) as the number of constraints is insufficient.

Since our study assumes that there is only one target speaker, we refer to the case when multiple microphones are available as \textit{over-determined condition}, following the conventions in source separation \cite{Sawada2019BSSReview}.
}

\subsection{Overview of USDnet}

\ZQHL{With the problem formulation in Section \ref{formulation_as_blind_deconv} and understandings in \ref{over-determined_disucussion}}, in the next section, following UNSSOR \cite{Wang2023UNSSOR} we propose USDnet, which tackles this problem by modeling speech patterns (i.e., source prior) via unsupervised deep learning.
The high-level idea is to leverage unsupervised deep learning to first produce a source estimate.
With the source estimated in (\ref{ideal_loss}), filter estimation then becomes a simple linear regression problem, which has a closed-form solution.
Finally, an objective similar to (\ref{ideal_loss}), \ZQHL{ideally computed based on multiple microphones}, can be designed to regularize the source estimate to have it approximate $S_q$.

\section{USDnet}\label{system_overview_section}

\begin{figure}
  \centering  
  \includegraphics[width=8.5cm]{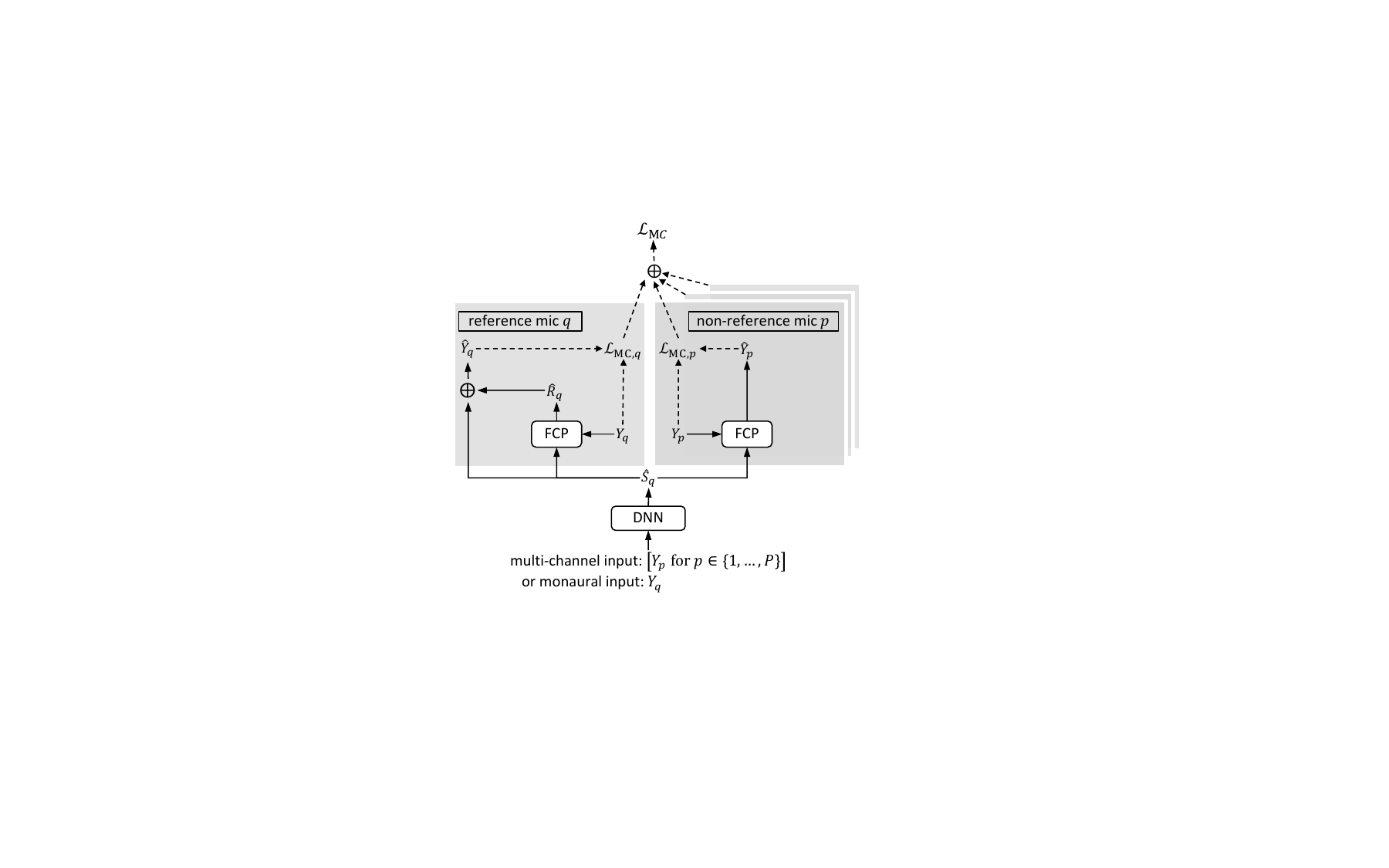}
  \caption{
  Illustration of \ZQHL{vanilla} USDnet (described in first paragraph of Section \ref{system_overview_section}).
  \ZQHL{Several variants are described in Section \ref{system_overview_section}.}
  }
  \label{system_figure}
\end{figure}

Fig. \ref{system_figure} illustrates the proposed USDnet \ZQHL{in its vanilla case}.
It takes as input the mixtures at all the microphones or just at the reference microphone $q$, and produces an estimate $\hat{S}_q$ at the reference microphone $q$.
To encourage $\hat{S}_q$ to approximate $S_q$, we regularize it during training by linearly filtering it via FCP \cite{Wang2021FCPjournal} so that the filtered estimates can approximate the mixtures at different microphones.
This section describes the DNN setup, loss functions, FCP filtering, an extension to monaural unsupervised dereverberation, and the rationale of various design choices of USDnet.

\subsection{DNN Setup}\label{DNN_configuration}

The DNN estimate $\hat{S}_q$ can be produced via complex spectral mapping \cite{Wang2020aCSMCHiME4, Wang2020css}, where the real and imaginary (RI) parts of input mixtures are stacked as input features for the DNN to predict the RI parts of $\hat{S}_q$.
Alternatively, it can be estimated via complex ratio masking \cite{Williamson2016, Williamson2017}, where the RI parts of a complex mask $\hat{M}_q$ is predicted by the DNN and $\hat{S}_q$ is obtained by $\hat{S}_q=\hat{M}_q \odot Y_q$, with $\odot$ denoting point-wise multiplication.
In supervised speech separation \cite{WDLreview}, both masking and mapping are popular, and mapping usually produces better separation than masking \cite{Wang2022GridNetjournal}.
In unsupervised dereverberation, however, we find that, for non-trivial reasons, training USDnet with masking is clearly better.
We will discuss the reasons in Section \ref{time_alignment}.
Other DNN configurations such as network architectures are provided later in Section \ref{mis_setup}.

\subsection{Mixture-Constraint Loss}\label{description_MC_loss}

Following UNSSOR \cite{Wang2023UNSSOR}, we design a mixture-constraint (MC) loss $\mathcal{L}_{\text{MC}}$ to regularize the DNN estimate $\hat{S}_q$ so that it can approximate $S_q$:
\begin{align}\label{MC_loss}
\mathcal{L}_{\text{MC}} = \mathcal{L}_{\text{MC},q} + \alpha \times \sum_{p=1, p\neq q}^{P} \mathcal{L}_{\text{MC},p},
\end{align}
with $\alpha \in \RR_{>0}$ denoting a microphone weighting term for non-reference microphones, and $\mathcal{L}_{\text{MC},q}$ and $\mathcal{L}_{\text{MC},p}$, defined by closely following the two terms in (\ref{ideal_loss}), as 
\begin{align}
\mathcal{L}_{\text{MC},q} &= \sum_{t,f} \mathcal{F} \Big( Y_q(t,f), \hat{Y}_q(t,f) \Big) \nonumber \\
&=\sum_{t,f} \mathcal{F} \Big( Y_q(t,f), \hat{S}_q(t,f) + \hat{R}_q(t,f) \Big) \nonumber \\
&=\sum_{t,f} \mathcal{F} \Big( Y_q(t,f), \hat{S}_q(t,f) + \hat{\mathbf{g}}_q(f)^\H \widetilde{\hat{\mathbf{S}}}_q(t,f) \Big), \label{MC_loss_ref} \\
\mathcal{L}_{\text{MC},p} &= \sum_{t,f} \mathcal{F}\Big( Y_p(t,f), \hat{Y}_p(t,f) \Big)\nonumber \\
&=\sum_{t,f} \mathcal{F}\Big( Y_p(t,f), \hat{\mathbf{h}}_p(f)^\H \overline{\hat{\mathbf{S}}}_q(t,f) \Big), \label{MC_loss_nonref}
\end{align}
where $\widetilde{\hat{\mathbf{S}}}_q(t,f) = \big[ \hat{S}_q(t-K+1,f),\dots,\hat{S}_q(t-\Delta,f)\big]^\T \in \CC^{K-\Delta}$ stacks a window of T-F units in $\hat{S}_q$ with a non-negative prediction delay $\Delta$, $\overline{\hat{\mathbf{S}}}_q(t,f) = \big[ \hat{S}_q(t-I+1,f),\dots,\hat{S}_q(t+J,f)\big]^\T \in \CC^{I+J}$, and $\hat{\mathbf{g}}_q(f)\in \CC^{K-\Delta}$ and $\hat{\mathbf{h}}_p(f)\in \CC^{I+J}$ respectively denote the corresponding estimated linear filters (to be described later).
\ZQHL{Notice that the filter taps, $K$, $\Delta$, $I$ and $J$, are unknown in unsupervised setups.
They are hyper-parameters to tune.
Their values are shared by all the training utterances in this study.}

Following \cite{Wang2021compensation, Wang2023UNSSOR}, $\mathcal{F}(\cdot, \cdot)$ computes a loss on the estimated RI components and their magnitude:
\begin{align}\label{L_D}
\mathcal{F} \Big( Y_a(t,f), \hat{Y}_a(t,f) \Big) = \nonumber \\
\frac{1}{\sum\nolimits_{t',f'} |Y_a(t',f')|} &\Big( \Big| \mathcal{R}(Y_a(t,f)) - \mathcal{R}(\hat{Y}_a(t,f)) \Big| \nonumber \\
&+ \Big| \mathcal{I}(Y_a(t,f)) - \mathcal{I}(\hat{Y}_a(t,f)) \Big| \nonumber \\
&+ \Big| |Y_a(t,f)| - |\hat{Y}_a(t,f)| \Big| \Big),
\end{align}
where $a$ indexes all the microphones, $|\cdot|$ computes magnitude, $\mathcal{R}(\cdot)$ and $\mathcal{I}(\cdot)$ respectively extract RI components, and the normalization term balances the losses at different microphones and across training mixtures.

To compute $\mathcal{L}_{\text{MC}}$ and enforce the above-mentioned regularizations to $\hat{S}_q$, we need to first estimate the filters, $\hat{\mathbf{g}}_q(f)$ and $\hat{\mathbf{h}}_p(f)$.
We describe the method in the next subsection.

In our experiments, we observe that minimizing $\mathcal{L}_{\text{MC}}$ can promote unsupervised dereverberation.
A plot of its loss curve is described later in Section \ref{loss_surface} and shown later in Fig. \ref{loss_surface_figure}.

\ZQHL{
Notice that we want $\hat{S}_q$ to approximate direct-path $S_q$, and $\hat{\mathbf{g}}_q(f)^\H \widetilde{\hat{\mathbf{S}}}_q(t,f)$ to approximate reverberation $R_q(t,f)$.
To achieve this, in (\ref{MC_loss_ref}) we constrain their summation to approximate the mixture (i.e., $\hat{Y}_q(t,f)=\hat{S}_q(t,f) + \hat{R}_q(t,f)=\hat{S}_q(t,f) + \hat{\mathbf{g}}_q(f)^\H \widetilde{\hat{\mathbf{S}}}_q(t,f)$), and in (\ref{MC_loss_nonref}) we constrain $\hat{S}_q$ such that it can be linearly filtered to approximate the mixture at another microphone (i.e., $\hat{Y}_p(t,f)=\hat{\mathbf{h}}_p(f)^\H \overline{\hat{\mathbf{S}}}_q(t,f)$).
However, naively using (\ref{MC_loss_ref}) and (\ref{MC_loss_nonref}) alone is not sufficiently effective at leading $\hat{S}_q$ to approximate $S_q$, as the loss is defined on the reconstructed mixture rather than on $\hat{S}_q$ itself.
We will propose more regularizations later in Section \ref{fcp_description_alternative}, \ref{Necessity_over_determined}, \ref{time_alignment} and \ref{why_approximating_directpath} so that even if the loss is defined on the reconstructed mixtures, the resulting $\hat{S}_q$ would still approximate $S_q$.
}

\subsection{FCP for Filter Estimation}\label{fcp_description}

Following UNSSOR \cite{Wang2023UNSSOR}, we compute the relative transfer functions $\hat{\mathbf{g}}_q(f)$ and $\hat{\mathbf{h}}_p(f)$ via forward convolutive prediction (FCP) \cite{Wang2021FCPjournal, Wang2021FCPwaspaa}, a neural forward filtering algorithm that linearly filters the DNN estimate $\hat{S}_q$ to approximate the mixture:
\begin{align}
&
\hat{\mathbf{g}}_q(f)
= \underset{\mathbf{g}_q(f)}{{\text{argmin}}} \sum\limits_t \frac{ \Big| Y_q(t,f)- \hat{S}_q(t,f)  - \mathbf{g}_q(f)^{\H}\ \widetilde{\hat{\mathbf{S}}}_q(t,f) \Big|^2}{\hat{\lambda}_q(t,f)}, 
\label{fcp_proj_mixture_dereverb_ref_mic} \\
&
\hat{\mathbf{h}}_p(f) = \underset{\mathbf{h}_p(f)}{{\text{argmin}}} \sum_t \frac{ \Big| Y_p(t,f) - \mathbf{h}_p(f)^{\H}\ \overline{\hat{\mathbf{S}}}_q(t,f) \Big|^2}{\hat{\lambda}_p(t,f)}, \label{fcp_proj_mixture_dereverb_nonref_mic}
\end{align}
where $\hat{\lambda}_a$, with $a$ indexing all the microphones, is a weighting term balancing the importance of various T-F units.
Following \cite{Wang2021FCPjournal, Wang2021FCPwaspaa, Wang2023UNSSOR}, we define it as 
\begin{align}\label{lambda}
\hat{\lambda}_a(t,f) = \frac{1}{P}\sum_{p=1}^P|Y_{p}(t,f)|^2 + \xi\times \text{max}\big(\frac{1}{P}\sum_{p=1}^P |Y_{p}|^2\big),
\end{align}
where $\xi$ (tuned to $10^{-4}$ in this paper) floors the weighting term to avoid placing too much weights on low-energy T-F units, and $\text{max}(\cdot)$ extracts the maximum value of a spectrogram.

Both (\ref{fcp_proj_mixture_dereverb_ref_mic}) and (\ref{fcp_proj_mixture_dereverb_nonref_mic}) are weighted linear regression problems, where closed-form solutions can be readily computed.
\ZQHL{For (\ref{fcp_proj_mixture_dereverb_nonref_mic}), the closed-form solution is
\begin{align}%
\hat{\mathbf{h}}_p(f) &= \nonumber \\
&\Big( \sum\limits_t \frac{\overline{\hat{\mathbf{S}}}_q(t,f) \overline{\hat{\mathbf{S}}}_q(t,f)^{\H}}{\hat{\lambda}_p(t,f)} \Big)^{-1} \sum\limits_t \frac{\overline{\hat{\mathbf{S}}}_q(t,f) \big( Y_p(t,f) \big)^{*}}{\hat{\lambda}_p(t,f)},\nonumber
\end{align}
where $(\cdot)^{*}$ computes complex conjugate.
For (\ref{fcp_proj_mixture_dereverb_ref_mic}), it is
\begin{align}%
&\hat{\mathbf{g}}_q(f) = \nonumber \\
&\Big( \sum\limits_t \frac{\widetilde{\hat{\mathbf{S}}}_q(t,f) \widetilde{\hat{\mathbf{S}}}_q(t,f)^{\H}}{\hat{\lambda}_q(t,f)} \Big)^{-1} \sum\limits_t \frac{\widetilde{\hat{\mathbf{S}}}_q(t,f) \big( Y_q(t,f) - \hat{S}_q(t,f) \big)^{*}}{\hat{\lambda}_q(t,f)}.\nonumber
\end{align}}
We then plug the closed-form solutions to (\ref{MC_loss_ref}) and (\ref{MC_loss_nonref}), compute the $\mathcal{L}_{\text{MC}}$ loss in (\ref{MC_loss}), and optimize the DNN.

\subsection{Interpretation of MC Loss and FCP Filtering}\label{loss_interpretation}

To facilitate understanding, we give an interpretation of the proposed $\mathcal{L}_{\text{MC}}$ loss.

Suppose that $\mathcal{F}(\cdot, \cdot)$ is defined as $\mathcal{F} \big( Y_r(t,f), \hat{Y}_r(t,f) \big) = | Y_r(t,f) - \hat{Y}_r(t,f) |^2$ (rather than as the one in (\ref{L_D}), no weighting is used in FCP (i.e., $\hat{\lambda}_a(t,f)=1$), and $\alpha=1$, $\mathcal{L}_{\text{MC}}$ in (\ref{MC_loss}) can be alternatively formulated as 
\begin{align}\label{loss_filter_all_interpretation}
&\mathcal{L}_{\text{M}}{}_{\text{C,alt}} = \nonumber \\
&\sum_{f} \underset{\mathbf{g}_q(f)}{{\text{min}}} \sum_{t} \Big| Y_q(t,f) - \hat{S}_q(t,f) - \mathbf{g}_q(f)^\H \widetilde{\hat{\mathbf{S}}}_q(t,f) \Big|^2 \nonumber \\
&+\sum_{p=1, p\neq q}^{P} \sum_{f} \underset{\mathbf{h}_p(f)}{{\text{min}}} \sum_{t} \Big| Y_p(t,f) - \mathbf{h}_p(f)^\H \overline{\hat{\mathbf{S}}}_q(t,f) \Big|^2,
\end{align}
which, given a DNN estimate $\hat{S}_q$, first computes the relative transfer functions $\hat{\mathbf{g}}_q(f)$ and $\hat{\mathbf{h}}_p(f)$ that best relate the DNN estimate $\hat{S}_q$ to the mixtures (i.e., best satisfy the physical mixture constraints) by minimizing a quadratic objective per frequency at each microphone, and then uses the minimum as the loss for DNN training.
The proposed approach can be considered as a \textit{minimizing the minimum} approach.
This echoes the idea of permutation invariant training \cite{Kolbak2017} in supervised speech separation, where the minimum to minimize is first computed by finding the best speaker permutation and then used as the loss to train DNNs.

In our study, we use $\mathcal{L}_{\text{MC}}$ in (\ref{MC_loss}) rather than $\mathcal{L}_{\text{MC,alt}}$ for DNN training, as we find that using $\mathcal{F}(\cdot, \cdot)$ defined in (\ref{L_D}) and the weighting term in (\ref{lambda}) produces better performance.

\subsection{Alternative FCP Filtering at Reference Microphone}\label{fcp_description_alternative}

An alternative way to compute $\hat{\mathbf{g}}_q(f)$ is as follows:
\begin{align}
\hat{\mathbf{g}}_q(f)
= \underset{\mathbf{g}_q(f)}{{\text{argmin}}} \sum\limits_t \frac{ \Big| Y_q(t,f)  - \mathbf{g}_q(f)^{\H}\ \widetilde{\hat{\mathbf{S}}}_q(t,f) \Big|^2}{\hat{\lambda}_q(t,f)},
\label{fcp_proj_mixture_dereverb_ref_mic_alternative}
\end{align}
where the difference from (\ref{fcp_proj_mixture_dereverb_ref_mic}) is that, in the numerator, $\hat{S}_q$ is not subtracted from $Y_q$.
Although this change appears minor, its benefits are non-trivial.
We find it \ZQHL{a useful regularization that can} (a) help USDnet avoid the trivial solution of copying the input mixture as the output (i.e., $\hat{S}_q=Y_q$); (b) enable USDnet to be trained on monaural mixtures \ZQHL{where only one microphone can be used in loss computation}; and (c) lead to an $\hat{S}_q$ that better approximates $S_q$ and a $\hat{\mathbf{g}}_q(f)^{\H}\ \widetilde{\hat{\mathbf{S}}}_q(t,f)$ that better approximates $R_q$, rather than an $\hat{S}_q$ and a $\hat{\mathbf{g}}_q(f)^{\H}\ \widetilde{\hat{\mathbf{S}}}_q(t,f)$ whose summation in (\ref{MC_loss_ref}) can better approximate the mixture $Y_q$ (and hence better minimize $\mathcal{L}_{\text{MC},q}$) but not respectively better approximating $S_q$ and $R_q$.
\ZQHL{We will later detail the first two benefits in Section \ref{Necessity_over_determined} and the third in \ref{why_approximating_directpath}, after describing the input to USDnet in the next subsection.}

\ZQHL{
\subsection{Number of Microphones in Input and Loss}\label{monaural_dereverb}

 We can use a subset of microphones in the input or loss to train USDnet.
 This section describes three cases.

\subsubsection{Multi-channel input and loss}

So far, we have mainly presented the loss functions of USDnet, which are computed based on multiple microphones.
On the input side, we can feed all the microphone mixtures (i.e., same as the ones used in loss computation) to USDnet.
The resulting trained USDnet performs multi-channel unsupervised dereverberation at run time.

\subsubsection{Single-channel input and multi-channel loss}

USDnet can be trained to realize monaural dereverberation at run time by only feeding in the mixture at the reference microphone (i.e., $Y_q$) to the DNN while still optimizing $\mathcal{L}_{\text{MC}}$ computed based on multi-channel mixtures.
Fig. \ref{system_figure} shows the idea.
The loss computed on multiple microphones could guide USDnet to model the spectro-temporal patterns in reverberant speech and realize monaural unsupervised dereverberation.
Notice that multi-channel mixtures are only required for training.
At run time, the trained model performs monaural dereverberation.

\subsubsection{Single-channel input and loss}

If the training mixtures are all monaural, we find that we can still train USDnet (although with very limited improvement) if the FCP filtering in (\ref{fcp_proj_mixture_dereverb_ref_mic_alternative}) is used, but the training would fail if (\ref{fcp_proj_mixture_dereverb_ref_mic}) is used for filter estimation.
We will describe the rationale in Section \ref{Necessity_over_determined}.
}

\ZQHL{\subsection{Necessity of Over-determined Training Mixtures}\label{Necessity_over_determined}

This subsection discusses the necessity of over-determined training mixtures, depending on whether $\hat{\mathbf{g}}_q(f)$ is computed using (\ref{fcp_proj_mixture_dereverb_ref_mic}) or (\ref{fcp_proj_mixture_dereverb_ref_mic_alternative}).

\subsubsection{When $\hat{\mathbf{g}}_q(f)$ is computed using (\ref{fcp_proj_mixture_dereverb_ref_mic})}

In this case, the mixtures for training USDnet need to be over-determined.
If the training mixtures are instead all monaural (i.e., using single-channel input and loss), $\mathcal{L}_{\text{MC}}$ would only contain the first term $\mathcal{L}_{\text{MC},q}$, which can be optimized to zero by the DNN by simply copying the input mixture as the output (i.e., $\hat{S}_q=Y_q$ and, in this case, $\hat{\mathbf{g}}_q(f)$ computed via (\ref{fcp_proj_mixture_dereverb_ref_mic}) would be all-zero since in the numerator $Y_q - \hat{S}_q=0$).
This trivial solution would not yield any dereverberation.
By leveraging extra microphone recordings as constraints, the second term in $\mathcal{L}_{\text{MC}}$ defined in (\ref{MC_loss}) could help avoid the trivial solution.

\subsubsection{When $\hat{\mathbf{g}}_q(f)$ is computed using (\ref{fcp_proj_mixture_dereverb_ref_mic_alternative})}

In this case, we find that the mixtures for training USDnet can also be monaural (i.e., by using single-channel input and loss).
Different from the previous case, in this case if the DNN simply copies the input as output (i.e., $\hat{S}_q=Y_q$), (\ref{fcp_proj_mixture_dereverb_ref_mic_alternative}) becomes identical to the optimization problem in WPE \cite{Nakatani2010}, a classic dereverberation algorithm.
Following WPE, the filtering result $\hat{\mathbf{g}}_q(f)^{\H}\ \widetilde{\hat{\mathbf{S}}}_q(t,f)=\hat{\mathbf{g}}_q(f)^{\H}\ \widetilde{Y}_q(t,f)$ would inevitably, and largely only, approximate late reverberation if the prediction delay $\Delta$ used in defining $\widetilde{\hat{\mathbf{S}}}_q(t,f)$ is reasonably large\footnote{This is because when the prediction delay $\Delta$ is set reasonably large, the target speech in the current T-F unit becomes much less linearly-correlated with the T-F units that are beyond $\Delta$ frames in the past \cite{Nakatani2010}.}.
As a result, the reconstructed mixture used for computing $\mathcal{L}_{\text{MC},q}$ in (\ref{MC_loss_ref}), $\hat{Y}_q(t,f) = \hat{S}_q(t,f)+\hat{\mathbf{g}}_q(f)^{\H}\ \widetilde{\hat{\mathbf{S}}}_q(t,f)=Y_q(t,f)+\hat{\mathbf{g}}_q(f)^{\H}\ \widetilde{Y}_q(t,f)$, would produce a large loss, due to the approximated late reverberation $\hat{\mathbf{g}}_q(f)^{\H}\ \widetilde{Y}_q(t,f)$.
Therefore, using (\ref{fcp_proj_mixture_dereverb_ref_mic_alternative}) and minimizing $\mathcal{L}_{\text{MC},q}$ could help USDnet avoid the trivial solution.
For similar reasons, using (\ref{fcp_proj_mixture_dereverb_ref_mic_alternative}) could yield better dereverberation than (\ref{fcp_proj_mixture_dereverb_ref_mic}) when multiple microphones are used in loss computation.
}

\subsection{Difficulties of Time Alignment, and Masking vs. Mapping}\label{time_alignment}

In supervised speech separation, the availability of target speech signals can provide a strong supervision at each sample for DNNs to learn to produce an estimate that is strictly time-aligned with target speech.
In WPE, a time-aligned estimate can be obtained by only removing estimated late reverberation.
In unsupervised neural dereverberation, however, we find it particularly difficult to produce an estimate time-aligned with target speech, simply due to the lack of an explicit sample-level supervision (e.g., oracle target speech).
For $\mathcal{L}_{\text{MC},q}$ in (\ref{MC_loss_ref}), even if $\hat{S}_q$ produced by the DNN is not time-aligned with the oracle target speech $S_q$, $\mathcal{L}_{\text{MC},q}$ can still be small because the filter $\hat{\mathbf{g}}_q(f)$, which is multi-tap, could compensate the time-misalignment through multi-frame filtering.
This compensation can also happen when we compute $\mathcal{L}_{\text{MC},p}$ in (\ref{MC_loss_nonref}).

To partially address this issue, we propose to use complex masking instead of mapping to obtain target estimate $\hat{S}_q$, and observe that masking obtains better performance.
This could be because, in masking, $\hat{S}_q$ is produced by masking the mixture $Y_q$, while, in mapping, $\hat{S}_q$ is produced directly by the DNN and could hence more likely be misaligned with target speech.
For example, when using complex mapping, at each T-F unit the DNN could just output an estimated signal not time- and gain-aligned with the mixture, e.g., $\hat{S}_q(t,f) = \beta \times Y_q(t+\Delta,f)$ where $\beta \in \CC$ denotes an arbitrary scaling factor with $|\beta|$ being small.
In this case, in (\ref{fcp_proj_mixture_dereverb_ref_mic_alternative}), $\hat{\mathbf{g}}_q(f)^{\H}\ \widetilde{\hat{\mathbf{S}}}_q(t,f)$ can perfectly approximate $Y_q(t,f)$ since now $\widetilde{\hat{\mathbf{S}}}_q(t,f) = \big[ \hat{S}_q(t-K,f),\dots,\hat{S}_q(t-\Delta,f)\big]^\T = \beta \times \big[ Y_q(t+\Delta-K,f),\dots,Y_q(t,f)\big]^\T$ has $Y_q(t,f)$ in the last element; and in (\ref{MC_loss_ref}), the resulting reconstructed mixture $\hat{Y}_q(t,f) = \hat{S}_q(t,f) + \hat{\mathbf{g}}_q(f)^\H \widetilde{\hat{\mathbf{S}}}_q(t,f)=\beta \times Y_q(t+\Delta,f) + Y_q(t,f)$ would be very close to the mixture $Y_q(t,f)$ if $|\beta|$ is small, yielding a very small $\mathcal{L}_{\text{MC},q}$ even though the solution is trivial.
By using masking, this trivial solution could be avoided, \ZQHL{as masking applies a complex-valued filter to the mixture at each T-F unit and it is difficult to modify the phase and magnitude of the mixture in a way such that the resulting estimates would be multiple frames ahead or behind}.

\ZQHL{
\subsection{Discussion on Why $\hat{S}_q$ Could Approximate $S_q$}\label{why_approximating_directpath}

So far, we have described the benefits of over-determined conditions in Section \ref{over-determined_disucussion}, and various regularizations such as using (\ref{fcp_proj_mixture_dereverb_ref_mic_alternative}) rather than (\ref{fcp_proj_mixture_dereverb_ref_mic}) for FCP filtering and using masking rather than mapping to obtain target estimate $\hat{S}_q$.
One more important regularization we have not discussed is the non-negative prediction delay $\Delta$.
We find that these regularizations combined can lead $\hat{S}_q$ to approximate $S_q$. This subsection provides more intuitions and discussions.

In (\ref{MC_loss_ref}), although the loss is defined on the reconstructed mixture $\hat{Y}_q(t,f)=\hat{S}_q(t,f)+\hat{\mathbf{g}}_q(f)^\H \widetilde{\hat{\mathbf{S}}}_q(t,f)$, our goal is to have $\hat{S}_q$ approximate direct-path $S_q$ and $\hat{\mathbf{g}}_q(f)^\H \widetilde{\hat{\mathbf{S}}}_q(t,f)$ approximate reverberation $R_q(t,f)$.
In this case, the prediction delay $\Delta$ has to be positive.
If it is zero, $\hat{S}_q$ can be absorbed into $\hat{\mathbf{g}}_q(f)^\H \widetilde{\hat{\mathbf{S}}}_q(t,f)$, and (\ref{MC_loss_ref}) would have the same form as (\ref{MC_loss_nonref}).
This would not lead $\hat{S}_q$ to approximate $S_q$, as there are infinite solutions to the estimated filter $\hat{\mathbf{g}}_q$ and source $\hat{S}_q$ that can minimize the losses in (\ref{MC_loss_ref}) and (\ref{MC_loss_nonref}).

Suppose that the prediction delay $\Delta$ is positive, masking is used, and noise $\varepsilon$ is negligible.
The loss in (\ref{MC_loss_ref}) would encourage $\hat{S}_q$ to approximate $S_q$.
Our interpretation consists of three parts:
\begin{itemize}[leftmargin=*,noitemsep,topsep=0pt]
\item When masking is used, $\hat{S}_q$ would be constrained to be approximately time-aligned with $S_q$ (see our discussion in Section \ref{time_alignment}).
\item Due to the positive prediction delay, $\hat{\mathbf{g}}_q(f)^\H \widetilde{\hat{\mathbf{S}}}_q(t,f)$, with $\hat{\mathbf{g}}_q(f)$ computed via (\ref{fcp_proj_mixture_dereverb_ref_mic_alternative}), largely can only approximate late reverberation.
\item To minimize (\ref{MC_loss_ref}), $\hat{S}_q$ and $\hat{\mathbf{g}}_q(f)^\H \widetilde{\hat{\mathbf{S}}}_q(t,f)$ need to add up to the mixture.
Since $\hat{\mathbf{g}}_q(f)^\H \widetilde{\hat{\mathbf{S}}}_q(t,f)$ can only approximate (or \textit{explain}) the late reverberation component of the mixture, $\hat{S}_q$ has to \textit{explain} the rest components, which mainly include (a) the direct-path signal $S_q$; and (b) some early reflections that cannot be explained by $\hat{\mathbf{g}}_q(f)^\H \widetilde{\hat{\mathbf{S}}}_q(t,f)$ when the prediction delay $\Delta$ is set to a large positive value.
\end{itemize}

This interpretation highlights the importance of using masking together with a non-negative prediction delay $\Delta$.
If mapping is used instead, it could predict multiple frames ahead and break the regularizations enforced by the positive prediction delay.

The third point above suggests that $\hat{S}_q$ would inevitably contain some early reflections due to the positive prediction delay.
Similarly to that in WPE \cite{Nakatani2010}, this could be fine in many applications. For example, including some early reflections can improve speech intelligibility \cite{Moore2012Book}.

On the other hand, the third point also suggests that we may need a noise estimate (besides the speech estimate) to \textit{explain} the noise component in the mixture.
We will describe this later in Section \ref{garbage_description}.
}

\subsection{Input Microphone Dropout}

When multiple microphone signals are used as the network input and in loss computation, we find that USDnet often optimizes the loss well but the dereverberation result is poor.
This could be because USDnet sees, from the input, \ZQHL{exactly} the same mixtures used in loss computation and, therefore, could figure out a way to optimize the loss well \ZQHL{but not dereverberate well}.

To deal with this issue, we propose to apply dropout \cite{Tompson2015} to input non-reference microphone signals, where, during training, each non-reference microphone signal in the input is entirely dropped out with a probability of $\theta$.

\subsection{Introducing Garbage Sources}\label{garbage_description}

To model environmental noises \ZQHL{(assumed weak)} and the modeling error caused by narrow-band approximation (i.e., $\varepsilon'$ in (\ref{phymodel_freq_reference}) and (\ref{phymodel_freq_non_reference})), we propose to use a garbage source to absorb them.

We modify $\mathcal{L}_{\text{MC},q}$ in (\ref{MC_loss_ref}) and $\mathcal{L}_{\text{MC},p}$ in (\ref{MC_loss_nonref}) as follows:
\begin{align}
\mathcal{L}_{\text{MC},q} &= \sum_{t,f} \mathcal{F} \Big(Y_q(t,f), \nonumber \\
&\hat{S}_q(t,f) + \hat{\mathbf{g}}_q(f)^\H \widetilde{\hat{\mathbf{S}}}_q(t,f) + \hat{\mathbf{w}}_q(f)^\H \Breve{\hat{\mathbf{V}}}_q(t,f) \Big), \label{MC_loss_ref_garbage}
\end{align}
\begin{align}
\mathcal{L}_{\text{MC},p} = \sum_{t,f} \mathcal{F} \Big( &Y_p(t,f), \nonumber \\
&\hat{\mathbf{h}}_p(f)^\H \overline{\hat{\mathbf{S}}}_q(t,f) + \hat{\mathbf{w}}_p(f)^\H \Breve{\hat{\mathbf{V}}}_q(t,f) \Big), \label{MC_loss_nonref_garbage}
\end{align}
where the DNN is trained to output one more spectrogram $\hat{V}_q$ in addition to $\hat{S}_q$, $\Breve{\hat{\mathbf{V}}}_q(t,f) = \big[ \hat{V}_q(t-L,f),\dots,\hat{V}_q(t+L,f)\big]^\T\in \CC^{L+1+L}$ stacks a window of $L+1+L$ T-F units in $\hat{V}_q$, and $\hat{\mathbf{g}}_q(f)$ and $\hat{\mathbf{h}}_p(f)$ are respectively computed in the same ways as in (\ref{fcp_proj_mixture_dereverb_ref_mic}) or (\ref{fcp_proj_mixture_dereverb_ref_mic_alternative}), and (\ref{fcp_proj_mixture_dereverb_nonref_mic}).
$\hat{\mathbf{w}}_q(f)$ and $\hat{\mathbf{w}}_p(f)$ are computed as follows, similarly to (\ref{fcp_proj_mixture_dereverb_nonref_mic}):
\begin{align}
\hat{\mathbf{w}}_a(f) = \underset{\mathbf{w}_a(f)}{{\text{argmin}}} \sum_t \frac{ \Big| Y_a(t,f) - \mathbf{w}_a(f)^{\H}\ \Breve{\hat{\mathbf{V}}}_q(t,f) \Big|^2}{\hat{\lambda}_a(t,f)}, \label{fcp_proj_mixture_dereverb_garbage}
\end{align}
where $a$ indexes all the microphones.

We point out that using garbage sources to absorb modeling errors is a widely-adopted technique in conventional separation algorithms such as IVA \cite{Sawada2019BSSReview}, and CAGMM- and MESSL-based spatial clustering \cite{Mandel2010MESSL, Boeddeker2021}.
Our novelty here is adapting this idea for DNN-based unsupervised dereverberation.

\subsection{Run-Time Inference}\label{run_time_inference}

At run time, we use $\hat{S}_q$ as the final prediction.
We just need to run feed-forwarding for \ZQHL{inference}, without needing to compute FCP filters.

In other words, USDnet is trained to optimize the $\mathcal{L}_{\text{MC}}$ loss in (\ref{MC_loss}) so that it can learn to model signal patterns in reverberant speech for dereverberation.
At run time, it can be used to dereverberate signals in the same way as models trained in a supervised way.
This novel design makes USDnet possible to be readily integrated with semi-supervised learning, where, for example, (a) USDnet trained on massive unlabeled data can be fine-tuned on labeled target-domain data via supervised learning; and (b) pre-trained supervised models can be adapted in an unsupervised way on target-domain unlabeled data by initializing USDnet with the pre-trained model.

Following WPE \cite{Nakatani2010}, we can alternatively use $Y_q(t,f)  - \hat{\mathbf{g}}_q(f)^{\H}\ \widetilde{\hat{\mathbf{S}}}_q(t,f)$ as the final prediction, where the subtracted term is obtained by filtering past DNN estimates with a prediction delay $\Delta$ (in $\widetilde{\hat{\mathbf{S}}}_q(t,f)$) and hence can be viewed as an estimate of reverberation to remove.
However, in our experiments the performance is worse than directly using $\hat{S}_q$.

\section{Experimental Setup}\label{setup}

We train USDnet using an existing speech dereverberation dataset, and evaluate it on its simulated test set and the REVERB dataset.
This section describes the datasets, miscellaneous setup, baseline systems, and evaluation metrics.

\subsection{WSJ0CAM-DEREVERB Dataset}

WSJ0CAM-DEREVERB is simulated and has been used in several supervised dereverberation studies \cite{Wang2021LowDistortion, Wang2021FCPjournal, Wang2022GridNetjournal}, which reported strong results.
The dry source signals are from the WSJ0CAM corpus, which has $7,861$, $742$ and $1,088$ utterances respectively in its training, validation and test sets.
Based on them, $39,293$ ($\sim$$77.7$ h), $2,968$ ($\sim$$5.6$ h) and $3,262$ ($\sim$$6.4$ h) noisy-reverberant mixtures are respectively simulated as the training, validation and test sets.
For each utterance, a room with random room characteristics and speaker and microphone locations is sampled.
The simulated microphone array has $8$ microphones uniformly placed on a circle with a diameter of $20$ cm.
The speaker-to-array distance is drawn from the range $[0.75, 2.5]$ m and the reverberation time (T60) from $[0.2, 1.3]$ s.
For each utterance, an $8$-channel diffuse air-conditioning noise is sampled from the REVERB dataset \cite{Kinoshita2016}, and added to the reverberant speech at an SNR (between the direct-path signal and the noise) sampled from the range $[5, 25]$ dB.
The sampling rate is $16$ kHz.

\subsection{REVERB Dataset}

We evaluate USDnet, trained on WSJ0CAM-DEREVERB, directly on the REVERB dataset \cite{Kinoshita2016}, which has real near- and far-field reverberant speech recorded in rooms with air-conditioning noises by an 8-microphone circular array with a diameter of $20$ cm.
The rooms have a T60 at around $0.7$ s and the speaker to microphone distances are around $1$ m in the near-field case and $2.5$ m in the far-field case.
We feed dereverberated signals produced by USDnet to an ASR backend for recognition.
The ASR backend is trained by using the default recipe in Kaldi\footnote{commit 61637e6c8ab01d3b4c54a50d9b20781a0aa12a59}, which trains a TDNN-based acoustic model based on the default multi-condition training data consisting of the noisy-reverberant speech of REVERB.
The sampling rate is $16$ kHz.

We emphasize that the purpose of this evaluation is to show the effectiveness of USDnet on real-recorded data and compare its performance with WPE, not to obtain state-of-the-art ASR performance on REVERB.

\subsection{Miscellaneous System Setup}\label{mis_setup}

The STFT window size is $32$ ms, hop size $8$ ms, and the square root of Hann window is used as the analysis window.
We use $512$-point discrete Fourier transform to extract $257$-dimensional complex spectra at each frame.

We employ TF-GridNet \cite{Wang2022GridNetjournal} as the DNN architecture.
Using symbols defined in Table I of \cite{Wang2022GridNetjournal}, we set its hyper-parameters to $D=48$, $B=4$, $I=4$, $J=4$, $H=192$, $L=4$ and $E=2$.
Please do not confuse the symbols of TF-GridNet with the ones in this paper.
When the DNN is trained for complex ratio masking, we truncate the RI parts of the estimated masks into the range $[-5, 5]$ before multiplying it with the mixture.

We consider $1$-, $2$-, $4$- and $8$-microphone dereverberation.
The first microphone is always considered as the referenece microphone.
For $1$-channel dereverberation, we use the mixture signal captured by microphone $1$; for $2$-channel dereverberation, we use microphone $1$ and $4$; for $4$-channel dereverberation, we use microphone $1$, $3$, $5$ and $7$; and for $8$-channel dereverberation, all the $8$ microphones are used.

\subsection{Baseline Systems}\label{baselines}

We consider WPE \cite{Nakatani2010} as the baseline.
It is unsupervised and is so far the most popular and successful dereverberation algorithm.
We use the implementation in the \textit{nara\_wpe} toolkit \cite{Drude2020}.
The STFT configuration for WPE is the same as that in USDnet.
Following \cite{Drude2020, NakataniCB2019}, the filter tap is tuned to $37$ in monaural cases, to $10$ in $4$-channel cases, to $5$ in $8$-channel cases, the prediction delay is $3$, and $3$ iterations is performed.
\ZQHL{
Building upon the configurations of WPE, we additionally provide the results of DNN-WPE \cite{Kinoshita2017}, where the target power spectral density used in WPE is provided by a supervised dereverberation DNN.
The DNN has the same architecture as that in USDnet.
It is trained via complex masking in the same way as USDnet, but the loss is defined on the RI components and magnitude of estimated target speech $\hat{S}_q$ and is weighted by the summation of mixture magnitude, following the distance metric in (\ref{L_D}).
}

We considered comparing USDnet with unsupervised neural dereverberation models such as \cite{Bie2022, Saito2023DEREVERB, Fu2022MetricGANU}.
However, they only deal with monaural cases and require separate source prior models trained on anechoic speech or DNN-based metric models (such as DNSMOS) trained on annotated anechoic and reverberant speech, and cannot be directly trained solely on a set of reverberant signals.
Differently, USDnet exploits implicit regularizations afforded by multiple microphones and can be trained directly on a set of reverberant signals.
In other words, they utilize very different signal principles, which can be likely integrated with USDnet in future studies.
We therefore do not consider them as baselines for comparison, and mainly focus on showing the effectiveness of leveraging mixture constraints in this paper.

\subsection{Evaluation Metrics}\label{evaluation}

For evaluation on the test set of WSJ0CAM-DEREVERB, we use the direct-path signal at the first microphone (designated as the reference microphone) as the reference signal for metric computation.
We report scores of perceptual evaluation of speech quality (PESQ) \cite{Rix2001} and extended short-time objective intelligibility (eSTOI) \cite{Taal2011}.
They are widely-adopted objective metrics of speech quality and intelligibility.
We use the \textit{python-pesq} toolkit to report narrow-band MOS-LQO scores for PESQ, and the \textit{pystoi} toolkit for eSTOI.
Additionally, we report scale-invariant signal-to-distortion ratio (SI-SDR) \cite{LeRoux2018a} but do not consider it as a primary evaluation metric due to the difficulties, and in many applications unnecessities, of time alignment.
For the ASR evaluation on REVERB, we report word error rates (WER).

All of these metrics, except WER, favor estimated signals that are time-aligned with reference signals.

\section{Evaluation Results}\label{results}

This section validates USDnet and its design choices based on the WSJ0CAM-DEREVERB and REVERB datasets.
Table \ref{summary_hyperparam} lists the main hyper-parameters of USDnet.
We start with $I=40$, $J=0$, $K=40$, $\Delta=3$, no garbage sources (i.e., the filter tap $L$ is shown as ``-'' in the tables), no microphone weight (i.e., $\alpha=1$), and no dropout to input microphones (i.e., $\theta=0$).

\subsection{Effects of Using More Microphones in Loss and Input}

In row 2a of Table \ref{results_simulated_number_of_channels}, the training mixtures are all monaural.
The performance is not good, since only one microphone signal can be used as the mixture constraint and this is likely not enough for the DNN to solve the ill-posed problem and learn to figure out what the target signal is.
On the other hand, through the formulation of $\mathcal{L}_{\text{MC},q}$ in (\ref{MC_loss_ref}) and using (\ref{fcp_proj_mixture_dereverb_ref_mic_alternative}) for filter estimation, USDnet does not fail completely and shows some effectiveness.

In row 2b-2d of Table \ref{results_simulated_number_of_channels}, only one microphone signal is used in input but multiple microphone signals are used for loss computation.
We observe that the dereverberation performance gets much better, compared with row 2a.
This supports our idea that using extra microphone signals as mixture constraints in the loss can help USDnet better figure out the target signal.

In row 3b-3d, multiple microphone signals are used in input and all of them are used for loss computation.
We observe better performance than row 2a.

Next, we use the results in 2a-2d and 3b-3d as baselines, and validate the effects of various design choices in USDnet.

\begin{table}[t]
\scriptsize
\centering
\sisetup{table-format=2.2,round-mode=places,round-precision=2,table-number-alignment = center,detect-weight=true,detect-inline-weight=math}
\caption{\textsc{Main Hyper-Parameters of USDnet}}
\vspace{-0.1cm}
\label{summary_hyperparam}
\resizebox{1.0\columnwidth}{!}{
\setlength{\tabcolsep}{1pt}
\begin{tabular}{
cc
}
\toprule
Symbols & Description  \\
\midrule
$K$, $\Delta$ & $\widetilde{\hat{\mathbf{S}}}_q(t,f) = \big[ \hat{S}_q(t-K+1,f),\dots,\hat{S}_q(t-\Delta,f)\big]^\T \in \CC^{K-\Delta}$ \\
\midrule
$I$, $J$ & $\overline{\hat{\mathbf{S}}}_q(t,f) = \big[ \hat{S}_q(t-I+1,f),\dots,\hat{S}_q(t+J,f)\big]^\T \in \CC^{I+J}$ \\
\midrule
$L$ & $\Breve{\hat{\mathbf{V}}}_q(t,f) = \big[ \hat{V}_q(t-L,f),\dots,\hat{V}_q(t+L,f)\big]^\T\in \CC^{L+1+L}$ for garbage source \\
\midrule
$\alpha$ & Microphone weight for non-reference microphones \\
\midrule
$\theta$ & Dropout ratio for input non-reference microphones \\
\bottomrule
\end{tabular}\vspace{-0.4cm}
}
\end{table}

\begin{table}[t]
\scriptsize
\centering
\captionsetup{justification=centering}
\sisetup{table-format=2.2,round-mode=places,round-precision=2,table-number-alignment = center,detect-weight=true,detect-inline-weight=math}
\caption{\textsc{Results on Test Set of WSJ0CAM-DEREVERB using Various Number of Microphone Signals in Input and Loss}}
\vspace{-0.1cm}
\label{results_simulated_number_of_channels}
\resizebox{1.02\columnwidth}{!}{
\setlength{\tabcolsep}{2pt}
\begin{tabular}{
c %
c
c %
c %
c %
c %
S[table-format=1,round-precision=0] %
S[table-format=1,round-precision=0] %
S[table-format=1.2,round-precision=2]
S[table-format=1.3,round-precision=3]
S[table-format=2.1,round-precision=1]
}
\toprule
{\multirow{2}{*}{\rotatebox[origin=c]{270}{Row}}} & {\multirow{2}{*}{Systems}} & \#Mics in & {\multirow{2}{*}{$\hat{\mathbf{g}}_q(f)$}} & {\multirow{2}{*}{$I/J/K/\Delta/L$}} & Masking/ & {\multirow{2}{*}{$\alpha$}} & {\multirow{2}{*}{$\theta$}} & {\multirow{2}{*}{PESQ$\uparrow$}} & {\multirow{2}{*}{eSTOI$\uparrow$}} & {SI-SDR} \\
 &  & input/loss & & & Mapping & & & & & {(dB)$\uparrow$} \\

\midrule

1 & Mixture & 1 / - & - & - & - & {-} & {-} & 1.6406105433337752 & 0.493757274092272 & -3.563633992835417 \\

\midrule

2a & USDnet & 1 / 1 & (\ref{fcp_proj_mixture_dereverb_ref_mic_alternative}) & 40 / 0 / 40 / 3 / - & Masking & 1 & {-} & 1.759163199430006 & 0.561225480546709 & -2.1473806877989357 \\
2b & USDnet & 1 / 2 & (\ref{fcp_proj_mixture_dereverb_ref_mic_alternative}) & 40 / 0 / 40 / 3 / - & Masking & 1 & {-} & 2.339252534378971 & 0.7354876254240188 & 1.6171214726292724 \\
2c & USDnet & 1 / 4 & (\ref{fcp_proj_mixture_dereverb_ref_mic_alternative}) & 40 / 0 / 40 / 3 / - & Masking & 1 & {-} & 2.5066823440779213 & 0.76010317287698 & 2.5082555081208655 \\
2d & USDnet & 1 / 8 & (\ref{fcp_proj_mixture_dereverb_ref_mic_alternative}) & 40 / 0 / 40 / 3 / - & Masking & 1 & {-} & 2.350717771170258 & 0.7420943201511349 & 2.2303781667320615 \\

\midrule

3b & USDnet & 2 / 2 & (\ref{fcp_proj_mixture_dereverb_ref_mic_alternative}) & 40 / 0 / 40 / 3 / - & Masking & 1 & 0 & 2.169792887954022 & 0.6960953966114424 & 1.1131660882679442 \\

3c & USDnet & 4 / 4 & (\ref{fcp_proj_mixture_dereverb_ref_mic_alternative}) & 40 / 0 / 40 / 3 / - & Masking & 1 & 0 & 2.279547061019701 & 0.7158828710134627 & 1.598769641992428 \\

3d & USDnet & 8 / 8 & (\ref{fcp_proj_mixture_dereverb_ref_mic_alternative}) & 40 / 0 / 40 / 3 / - & Masking & 1 & 0 & 2.330097478675667 & 0.7296095164191431 & 2.0467920889483717 \\

\midrule

4a & USDnet & 1 / 1 & (\ref{fcp_proj_mixture_dereverb_ref_mic_alternative}) & 40 / 0 / 40 / 3 / - & Mapping & 1 & {-} & 1.612181271825518 & 0.493112388789571 & -3.5600012302525412 \\

4b & USDnet & 1 / 8 & (\ref{fcp_proj_mixture_dereverb_ref_mic_alternative}) & 40 / 0 / 40 / 3 / - & Mapping & 1 & {-} & 1.8874909031033005 & 0.27939043617312537 & -35.958197095290345 \\

4c & USDnet & 8 / 8 & (\ref{fcp_proj_mixture_dereverb_ref_mic_alternative}) & 40 / 0 / 40 / 3 / - & Mapping & 1 & 0 & 1.8688184143501436 & 0.507280871312554 & -28.5158822724309 \\

\midrule

5a & USDnet & 1 / 1 & (\ref{fcp_proj_mixture_dereverb_ref_mic}) & 40 / 0 / 40 / 3 / - & Masking & 1 & {-} & 1.6405564406983768 & 0.4937556821767547 & -3.5637899052043562 \\

5b & USDnet & 1 / 8 & (\ref{fcp_proj_mixture_dereverb_ref_mic}) & 40 / 0 / 40 / 3 / - & Masking & 1 & {-} & 2.0768934065107185 & 0.6713192532881885 & -0.10374395160663945 \\

5c & USDnet & 8 / 8 & (\ref{fcp_proj_mixture_dereverb_ref_mic}) & 40 / 0 / 40 / 3 / - & Masking & 1 & 0 & 2.084950191560952 & 0.6776848027070137 & -0.02940932188959877 \\

\bottomrule
\end{tabular}%
}
\end{table}

\subsection{Illustration of Loss Curve of $\mathcal{L}_{\text{MC}}$}\label{loss_surface}

\begin{figure}
  \centering  
  \includegraphics[width=8.5cm]{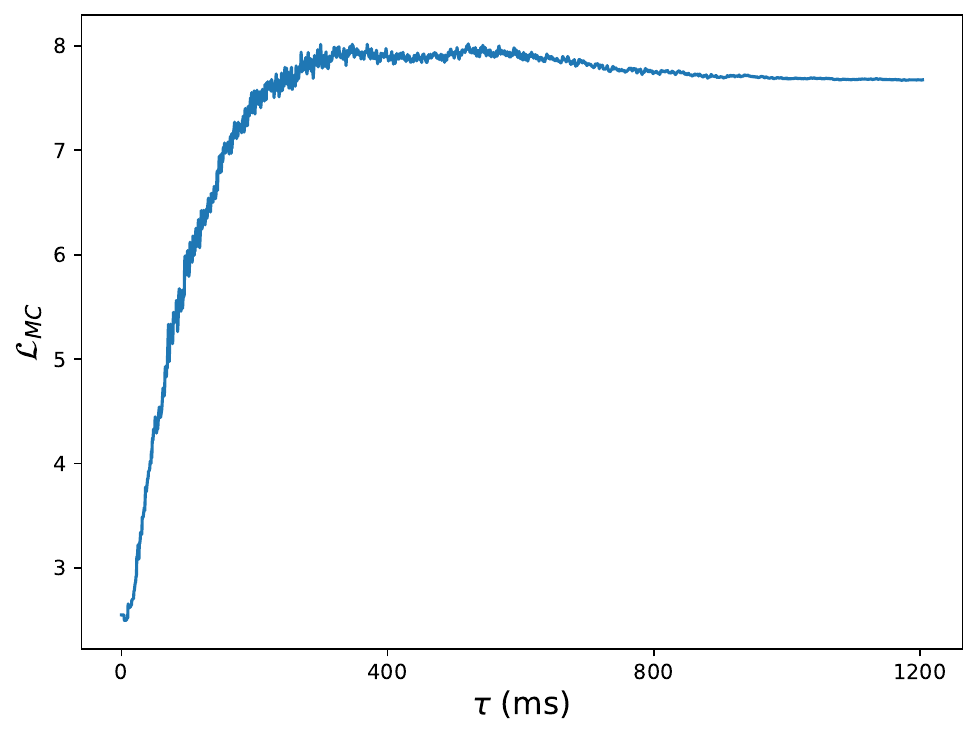}
  \vspace{-0.1cm}
  \caption{
  Loss curve of $\mathcal{L}_{\text{MC}}$ against truncated length of relative RIR, $\tau$. For this example utterance, the T60 is $1.204$ seconds.
  The truncation length is enumerated every $10$ samples, starting from one sample long all the way up to the length of T60.
  See Section \ref{loss_surface} for more details.
  }
  \label{loss_surface_figure}
\end{figure}

To show that minimizing $\mathcal{L}_{\text{MC}}$ can promote unsupervised dereverberation, we plot the loss curve of $\mathcal{L}_{\text{MC}}$ in Fig. \ref{loss_surface_figure}, by using a simulated noisy-reverberant speech signal sampled from WSJ0CAM-DEREVERB.
Given the simulated direct-path RIR $o_d\in \RR^{M_d}$ and reverberant-speech RIR $o_i\in \RR^{M_i}$, both at the reference microphone $q$, we first compute the relative RIR $o_r$ relating the direct-path signal to reverberant speech as follows:
\begin{align}
o_r = \text{iFFT}\Big(\frac{\text{FFT}(o_i, M_r)}{\text{FFT}(o_d, M_r)}, M_r\Big)\in \RR^{M_r},
\end{align}
where $M_r=M_i+M_d-1$, $\text{FFT}(\cdot, M_r)$ computes an $M_r$-point fast Fourier transform (FFT) of the RIR, and $\text{iFFT}(\cdot, M_r)$ computes an $M_r$-point inverse FFT (iFFT).
We then compute a hypothesized dereverberation result as
\begin{align}
\hat{S}_q = \text{STFT}\big(s_q * \operatorname{T}_{0}^{\tau}(o_r)\big),
\end{align}
where $s_q$ is the time-domain direct-path signal corresponding to $S_q$, the operator $*$ denotes linear convolution, and $\operatorname{T}_{0}^{\tau}(\cdot)$ truncates the RIR to length $\tau$ so that we can control how reverberant $\hat{S}_q$ is.
We then use $\hat{S}_q$ to compute $\mathcal{L}_{\text{MC}}$ based on the hyper-parameters shown in row 2d of Table \ref{results_simulated_number_of_channels}, and plot the loss curve against $\tau$.
From Fig. \ref{loss_surface_figure}, we observe much smaller $\mathcal{L}_{\text{MC}}$ when $\hat{S}_q$ is configured less reverberant.
This partially supports our claim that minimizing $\mathcal{L}_{\text{MC}}$ can promote dereverberation.

\subsection{Effects of Masking vs. Mapping}

In row 4a-4c of Table \ref{results_simulated_number_of_channels}, we use complex spectral mapping instead of complex masking.
The results are clearly worse than those in 2a, 2d and 3d, respectively, and, in particular, the SI-SDR scores are much lower.
This is because, in complex spectral mapping, the DNN has fewer restrictions in producing target estimates, and this would result in trivial solutions where reverberation is not reduced (see our discussions in Section \ref{time_alignment}).
During training, we indeed observe that, when using mapping, the training loss often suddenly decreases to values very close to zero.

In the rest of this paper, we use masking in default.

\subsection{Effects of (\ref{fcp_proj_mixture_dereverb_ref_mic_alternative}) vs. (\ref{fcp_proj_mixture_dereverb_ref_mic}) for FCP Filtering at Reference Mic}

In row 5a-5c of Table \ref{results_simulated_number_of_channels}, we use (\ref{fcp_proj_mixture_dereverb_ref_mic}) instead of (\ref{fcp_proj_mixture_dereverb_ref_mic_alternative}) to compute the filter $\hat{\mathbf{g}}_q(f)$ at the reference microphone $q$.
The results are respectively much worse than the ones in row 2a, 2d and 3d, indicating the effectiveness of using (\ref{fcp_proj_mixture_dereverb_ref_mic_alternative}).

In 5a, no improvement is observed over the unprocessed mixture.
This is because the network produces a trivial solution by copying the input mixture as the output.
See the discussions in the first paragraph of Section \ref{Necessity_over_determined}.
Differently, in 2a, we observe some improvement over the mixture by using (\ref{fcp_proj_mixture_dereverb_ref_mic_alternative}) instead of (\ref{fcp_proj_mixture_dereverb_ref_mic}) for FCP filtering.
The rationale is provided in the second paragraph of Section \ref{Necessity_over_determined}.

In the rest of this paper, we use (\ref{fcp_proj_mixture_dereverb_ref_mic_alternative}) in default.

\subsection{Effects of Non-Negative Prediction Delay $\Delta$}

The prediction delay $\Delta$ in defining $\widetilde{\hat{\mathbf{S}}}_q(t,f)$ (see the text below (\ref{MC_loss_nonref})) is a hyper-parameter to tune.
Ideally, for $\mathcal{L}_{\text{MC},q}$ in (\ref{MC_loss_ref}), we want $\hat{S}_q(t,f)$ to approximate the direct-path signal and $\hat{\mathbf{g}}_q(f)^\H \widetilde{\hat{\mathbf{S}}}_q(t,f)$ to approximate all the reflections so that their summation can add up to the mixture $Y_q$.
When using (\ref{fcp_proj_mixture_dereverb_ref_mic_alternative}) to compute $\mathbf{g}_q(f)$, a large $\Delta$ would make $\mathbf{g}_q(f)^{\H}\ \widetilde{\hat{\mathbf{S}}}_q(t,f)$ in the numerator not capable of approximating early reflections, and a small $\Delta$ could result in a $\mathbf{g}_q(f)^{\H}\ \widetilde{\hat{\mathbf{S}}}_q(t,f)$ that cancels out target direct-path speech.
In Table \ref{results_simulated_delta}, we sweep $\Delta$ based on the set of $\{1,2,3,4\}$, and observe that setting $\Delta$ to $3$ or $2$ results in better performance.
However, the performance differences are not very large.

\subsection{Effects of Filter Length $K$ and $I$}

The filter length $K$ and $I$ in defining $\widetilde{\hat{\mathbf{S}}}_q(t,f)$ and $\overline{\hat{\mathbf{S}}}_q(t,f)$ (see the text below (\ref{MC_loss_nonref})) are hyper-parameters to tune.
We sweep them based on the set of $\{20,40,60,80\}$.
Given that the STFT window and hop sizes are respectively $32$ and $8$ ms, this set of filter taps roughly samples filter lengths in the range of $[0.2, 0.7]$ seconds.
From Table \ref{results_simulated_filter_length}, we observe that using $I=60$ together with $K=60$ produces the best performance, but the differences in performance are not very large.

\begin{table}[t]
\scriptsize
\centering
\captionsetup{justification=centering}
\sisetup{table-format=2.2,round-mode=places,round-precision=2,table-number-alignment = center,detect-weight=true,detect-inline-weight=math}
\caption{\textsc{Results on Test Set of WSJ0CAM-DEREVERB using Various Prediction Delays $\Delta$}}
\vspace{-0.1cm}
\label{results_simulated_delta}
\setlength{\tabcolsep}{2pt}
\resizebox{1.0\columnwidth}{!}{
\begin{tabular}{
c %
c
c %
c %
S[table-format=1,round-precision=0] %
S[table-format=1,round-precision=0] %
S[table-format=1.2,round-precision=2]
S[table-format=1.3,round-precision=3]
S[table-format=1.1,round-precision=1]
}
\toprule
{\multirow{2}{*}{Row}} & {\multirow{2}{*}{Systems}} & \#CH in & {\multirow{2}{*}{$I/J/K/\Delta/L$}} & {\multirow{2}{*}{$\alpha$}} & {\multirow{2}{*}{$\theta$}} & {\multirow{2}{*}{PESQ$\uparrow$}} & {\multirow{2}{*}{eSTOI$\uparrow$}} & {\multirow{2}{*}{SI-SDR (dB)$\uparrow$}} \\
 &  & input/loss & & & & & \\

\midrule

1 & USDnet & 8 / 8 & 40 / 0 / 40 / 1 / - & 1 & 0 & 2.3444066434678383 & 0.7166543993852691 & 2.3444066434678383 \\

2 & USDnet & 8 / 8 & 40 / 0 / 40 / 2 / - & 1 & 0 & 2.3400490119514257 & 0.7343376790164508 & 2.8202944036290805 \\

3 & USDnet & 8 / 8 & 40 / 0 / 40 / 3 / - & 1 & 0 & 2.330097478675667 & 0.7296095164191431 & 2.0467920889483717 \\

4 & USDnet & 8 / 8 & 40 / 0 / 40 / 4 / - & 1 & 0 & 2.324709243857736 & 0.7258627730883485 & 1.310272337295945 \\

\bottomrule
\end{tabular}%
}

\vspace{0.3cm}

\scriptsize
\centering
\captionsetup{justification=centering}
\sisetup{table-format=2.2,round-mode=places,round-precision=2,table-number-alignment = center,detect-weight=true,detect-inline-weight=math}
\caption{\textsc{Results on Test Set of WSJ0CAM-DEREVERB using Various Filter Length $K$ and $I$}}
\vspace{-0.1cm}
\label{results_simulated_filter_length}
\setlength{\tabcolsep}{2pt}
\begin{tabular}{
c %
c
c %
c %
S[table-format=1,round-precision=0] %
S[table-format=1,round-precision=0] %
S[table-format=1.2,round-precision=2]
S[table-format=1.3,round-precision=3]
S[table-format=1.1,round-precision=1]
}
\toprule
{\multirow{2}{*}{Row}} & {\multirow{2}{*}{Systems}} & \#CH in & {\multirow{2}{*}{$I/J/K/\Delta/L$}} & {\multirow{2}{*}{$\alpha$}} & {\multirow{2}{*}{$\theta$}} & {\multirow{2}{*}{PESQ$\uparrow$}} & {\multirow{2}{*}{eSTOI$\uparrow$}} & {\multirow{2}{*}{SI-SDR (dB)$\uparrow$}} \\
 &  & input/loss & & & & & \\

\midrule

1 & USDnet & 8 / 8 & 20 / 0 / 20 / 3 / - & 1 & 0 & 2.302729584189882 & 0.715440821757632 & 1.3222702851745256 \\

2 & USDnet & 8 / 8 & 40 / 0 / 40 / 3 / - & 1 & 0 & 2.330097478675667 & 0.7296095164191431 & 2.0467920889483717 \\

3 & USDnet & 8 / 8 & 60 / 0 / 60 / 3 / - & 1 & 0 & 2.353980152794074 & 0.7392999597322416 & 2.2848620760755054 \\

4 & USDnet & 8 / 8 & 80 / 0 / 80 / 3 / - & 1 & 0 & 2.267627137522987 & 0.7271824191065831 & 2.766748311463708 \\

\bottomrule
\end{tabular}%
\end{table}

\subsection{Effects of Filter Delay $J$ for Non-Reference Microphones}

In the text below (\ref{MC_loss_nonref}), $\widetilde{\hat{\mathbf{S}}}_q(t,f)$ is configured to only contain past T-F units, while $\overline{\hat{\mathbf{S}}}_q(t,f)$ is configured to have $J$ future T-F units.
In earlier experiments, $J$ is set to $0$, while, in this subsection, we increase it to $1$ to account for the fact that the reference microphone is not always the closest microphone to the target speaker.
Notice that we do not need more future taps, since the microphone array is a compact array with an aperture size of only $20$ cm, which is much small than the speed of sound.
From Table \ref{results_simulated_filter_delay}, \ZQHL{we observe worse performance in almost all the entries}.

\subsection{Effects of Weighting Microphones}

Table \ref{results_microphone_weight} presents the results of using a smaller microphone weight $\alpha$ in (\ref{MC_loss}) for non-reference microphones.
The motivation is that when setting $\alpha=1.0$ and many microphones are used to compute $\mathcal{L}_{\text{MC}}$ in (\ref{MC_loss}), the loss on non-reference microphones would dominate the overall loss, causing the loss on the reference microphone, $\mathcal{L}_{\text{MC},q}$, which is more relevant to the target signal we aim to estimate, not optimized well.
In our experiments, if there are $P$ ($>4$) microphones for loss computation, we set $\alpha=\frac{3}{P-1}$, and still set $\alpha=1.0$ otherwise.
In Table \ref{results_microphone_weight}, better performance is observed.

\begin{table}[t]
\scriptsize
\centering
\captionsetup{justification=centering}
\sisetup{table-format=2.2,round-mode=places,round-precision=2,table-number-alignment = center,detect-weight=true,detect-inline-weight=math}
\caption{\textsc{Results on Test Set of WSJ0CAM-DEREVERB using Various Filter Delay $J$}}
\vspace{-0.1cm}
\label{results_simulated_filter_delay}
\setlength{\tabcolsep}{2pt}
\resizebox{1.0\columnwidth}{!}{
\begin{tabular}{
c %
c
c %
c %
S[table-format=1,round-precision=0] %
S[table-format=1,round-precision=0] %
S[table-format=1.2,round-precision=2]
S[table-format=1.3,round-precision=3]
S[table-format=1.1,round-precision=1]
}
\toprule
{\multirow{2}{*}{Row}} & {\multirow{2}{*}{Systems}} & \#CH in & {\multirow{2}{*}{$I/J/K/\Delta/L$}} & {\multirow{2}{*}{$\alpha$}} & {\multirow{2}{*}{$\theta$}} & {\multirow{2}{*}{PESQ$\uparrow$}} & {\multirow{2}{*}{eSTOI$\uparrow$}} & {\multirow{2}{*}{SI-SDR (dB)$\uparrow$}} \\
 &  & input/loss & & & & & \\

\midrule

1 & USDnet & 1 / 8 & 40 / 0 / 40 / 3 / - & 1 & {-} & 2.350717771170258 & 0.7420943201511349 & 2.2303781667320615 \\

2 & USDnet & 1 / 8 & 39 / 1 / 40 / 3 / - & 1 & {-} & 2.355559416119852 & 0.7298711188555901 & 1.6395401093891264 \\

\midrule

3 & USDnet & 8 / 8 & 40 / 0 / 40 / 3 / - & 1 & 0 & 2.330097478675667 & 0.7296095164191431 & 2.0467920889483717 \\

4 & USDnet & 8 / 8 & 39 / 1 / 40 / 3 / - & 1 & 0 & 2.3143702580985814 & 0.719414782665487 & 1.5645292320188415 \\

\bottomrule
\end{tabular}%
}

\vspace{0.3cm}

\scriptsize
\centering
\captionsetup{justification=centering}
\sisetup{table-format=2.2,round-mode=places,round-precision=2,table-number-alignment = center,detect-weight=true,detect-inline-weight=math}
\caption{\textsc{Results on Test Set of WSJ0CAM-DEREVERB using Various Microphone Weights $\alpha$}}
\vspace{-0.1cm}
\label{results_microphone_weight}
\setlength{\tabcolsep}{2pt}
\resizebox{1.0\columnwidth}{!}{
\begin{tabular}{
c %
c
c %
c %
S[table-format=1,round-precision=0] %
S[table-format=1,round-precision=0] %
S[table-format=1.2,round-precision=2]
S[table-format=1.3,round-precision=3]
S[table-format=1.1,round-precision=1]
}
\toprule
{\multirow{2}{*}{Row}} & {\multirow{2}{*}{Systems}} & \#CH in & {\multirow{2}{*}{$I/J/K/\Delta/L$}} & {\multirow{2}{*}{$\alpha$}} & {\multirow{2}{*}{$\theta$}} & {\multirow{2}{*}{PESQ$\uparrow$}} & {\multirow{2}{*}{eSTOI$\uparrow$}} & {\multirow{2}{*}{SI-SDR (dB)$\uparrow$}} \\
 &  & input/loss & & & & & \\

\midrule

1 & USDnet & 1 / 8 & 40 / 0 / 40 / 3 / - & 1 & {-} & 2.350717771170258 & 0.7420943201511349 & 2.2303781667320615 \\

2 & USDnet & 1 / 8 & 40 / 0 / 40 / 3 / - & {$\frac{3}{8-1}$} & {-} & 2.4806705378155853 & 0.7485352414798231 & 2.2246829971965285 \\

\midrule

3 & USDnet & 8 / 8 & 40 / 0 / 40 / 3 / - & 1 & 0 & 2.330097478675667 & 0.7296095164191431 & 2.0467920889483717 \\

4 & USDnet & 8 / 8 & 40 / 0 / 40 / 3 / - & {$\frac{3}{8-1}$} & 0 & 2.398516595802155 & 0.7449368389728661 & 2.2450837646433173 \\

\bottomrule
\end{tabular}%
}

\vspace{0.3cm}

\scriptsize
\centering
\captionsetup{justification=centering}
\sisetup{table-format=2.2,round-mode=places,round-precision=2,table-number-alignment = center,detect-weight=true,detect-inline-weight=math}
\caption{\textsc{Results on Test Set of WSJ0CAM-DEREVERB using\\Input Microphone Dropout}}
\vspace{-0.1cm}
\label{results_input_microphone_dropout}
\setlength{\tabcolsep}{2pt}
\resizebox{1.0\columnwidth}{!}{
\begin{tabular}{
c %
c
c %
c %
S[table-format=1,round-precision=0] %
S[table-format=1.1,round-precision=1] %
S[table-format=1.2,round-precision=2]
S[table-format=1.3,round-precision=3]
S[table-format=1.1,round-precision=1]
}
\toprule
{\multirow{2}{*}{Row}} & {\multirow{2}{*}{Systems}} & \#CH in & {\multirow{2}{*}{$I/J/K/\Delta/L$}} & {\multirow{2}{*}{$\alpha$}} & {\multirow{2}{*}{$\theta$}} & {\multirow{2}{*}{PESQ$\uparrow$}} & {\multirow{2}{*}{eSTOI$\uparrow$}} & {\multirow{2}{*}{SI-SDR (dB)$\uparrow$}} \\
 &  & input/loss & & & & & \\

\midrule

1a & USDnet & 2 / 2 & 40 / 0 / 40 / 3 / - & 1 & 0 & 2.169792887954022 & 0.6960953966114424 & 1.1131660882679442 \\

1b & USDnet & 4 / 4 & 40 / 0 / 40 / 3 / - & 1 & 0 & 2.279547061019701 & 0.7158828710134627 & 1.598769641992428 \\

1c & USDnet & 8 / 8 & 40 / 0 / 40 / 3 / - & {$\frac{3}{8-1}$} & 0 & 2.398516595802155 & 0.7449368389728661 & 2.2450837646433173 \\

\midrule

2a & USDnet & 2 / 2 & 40 / 0 / 40 / 3 / - & 1 & 0.7 & 2.3716316692014914 & 0.7450083370004698 & 2.139048303784847 \\
2b & USDnet & 4 / 4 & 40 / 0 / 40 / 3 / - & 1 & 0.7 & 2.5331772501522747 & 0.7509520368010777 & 2.2960714362520394 \\
2c & USDnet & 8 / 8 & 40 / 0 / 40 / 3 / - & {$\frac{3}{8-1}$} & 0.7 & 2.4231708638511646 & 0.7462786091427627 & 2.1455779371709514 \\

\bottomrule
\end{tabular}%
}

\vspace{0.3cm}

\scriptsize
\centering
\captionsetup{justification=centering}
\sisetup{table-format=2.2,round-mode=places,round-precision=2,table-number-alignment = center,detect-weight=true,detect-inline-weight=math}
\caption{\textsc{Results on Test Set of WSJ0CAM-DEREVERB using\\Garbage Sources}}
\vspace{-0.1cm}
\label{results_garbage_sources}
\setlength{\tabcolsep}{2pt}
\resizebox{1.0\columnwidth}{!}{
\begin{tabular}{
c %
c
c %
c %
S[table-format=1,round-precision=0] %
S[table-format=1.1,round-precision=1] %
S[table-format=1.2,round-precision=2]
S[table-format=1.3,round-precision=3]
S[table-format=1.1,round-precision=1]
}
\toprule
{\multirow{2}{*}{Row}} & {\multirow{2}{*}{Systems}} & \#CH in & {\multirow{2}{*}{$I/J/K/\Delta/L$}} & {\multirow{2}{*}{$\alpha$}} & {\multirow{2}{*}{$\theta$}} & {\multirow{2}{*}{PESQ$\uparrow$}} & {\multirow{2}{*}{eSTOI$\uparrow$}} & {\multirow{2}{*}{SI-SDR (dB)$\uparrow$}} \\
 &  & input/loss & & & & & \\

\midrule

1 & Mixture & 1 / - & - & {-} & {-} & 1.6406105433337752 & 0.493757274092272 & -3.563633992835417 \\

\midrule

2a & USDnet & 1 / 8 & 60 / 0 / 60 / 3 / - & {$\frac{3}{8-1}$} & {-} & 2.4463804130682836 & 0.7569215634436278 & 2.495406880906926 \\
2b & USDnet & 8 / 8 & 60 / 0 / 60 / 3 / - & {$\frac{3}{8-1}$} & 0.7 & 2.4217523200098245 & 0.7595283343159086 & 2.5781748396445607 \\

\midrule

3a & USDnet & 1 / 8 & 60 / 0 / 60 / 3 / 1 & {$\frac{3}{8-1}$} & {-} & 2.5270067927005613 & 0.7716912101362126 & 2.8603948775779355 \\
3b & USDnet & 8 / 8 & 60 / 0 / 60 / 3 / 1 & {$\frac{3}{8-1}$} & 0.7 & 2.451282974111018 & 0.7605576527097211 & 2.675615118266085 \\ %

\midrule

4a & WPE \cite{Nakatani2010} & 1 / - & - & {-} & {-} & 1.777453876250657 & 0.5792489171414862 & -1.6990565286957235 \\
4b & WPE \cite{Nakatani2010} & 8 / - & - & {-} & {-} & 2.0220796733016546 & 0.6899447169423142 & 1.9984655773076692 \\

\midrule

5a & DNN-WPE \cite{Kinoshita2017} & 1 / - & - & {-} & {-} &  1.806205157046666 & 0.6110156636761548 & -1.2347896697806466 \\
5b & DNN-WPE \cite{Kinoshita2017} & 8 / - & - & {-} & {-} & 2.071846921410318 & 0.7240008832744722 & 2.7831842936820994 \\

\bottomrule
\end{tabular}%
}
\end{table}

\subsection{Effects of Input Microphone Dropout}

In Table \ref{results_input_microphone_dropout}, we report the results of applying input microphone dropout to input non-reference microphone signals during training.
The dropout probability $\theta$ of each non-reference microphone is $0.7$ in default.
Better performance is observed for all the $2$-, $4$- and $8$-channel cases.

\subsection{Effects of Using Garbage Sources}

In Table \ref{results_garbage_sources}, we increase the filter taps $I$ and $K$ to $60$ and report the results of using one garbage source to absorb environmental noises and modeling errors.
We tune the FCP filter for filtering the garbage source to $3$-tap, by setting $L$ to $1$ (see the text below (\ref{MC_loss_nonref_garbage}) for the definitions).
Comparing row 2a with 3a, and 2b with 3b, we observe noticeable improvements.

\subsection{Comparison with WPE and DNN-WPE on WSJ0CAM-DEREVERB}

\ZQHL{Table \ref{results_garbage_sources} also reports the results of monaural and eight-channel WPE and supervised DNN-WPE.
USDnet obtains clearly better performance than WPE, especially in monaural input cases (e.g., $2.53$ vs. $1.78$ in PESQ).
DNN-WPE obtains better results than WPE.
USDnet obtains better performance than DNN-WPE in almost all the entries, except on SI-SDR in the eight-channel-input case (i.e., $2.7$ vs. $2.8$ dB SI-SDR).}

\begin{figure}
  \centering  
  \includegraphics[width=8cm]{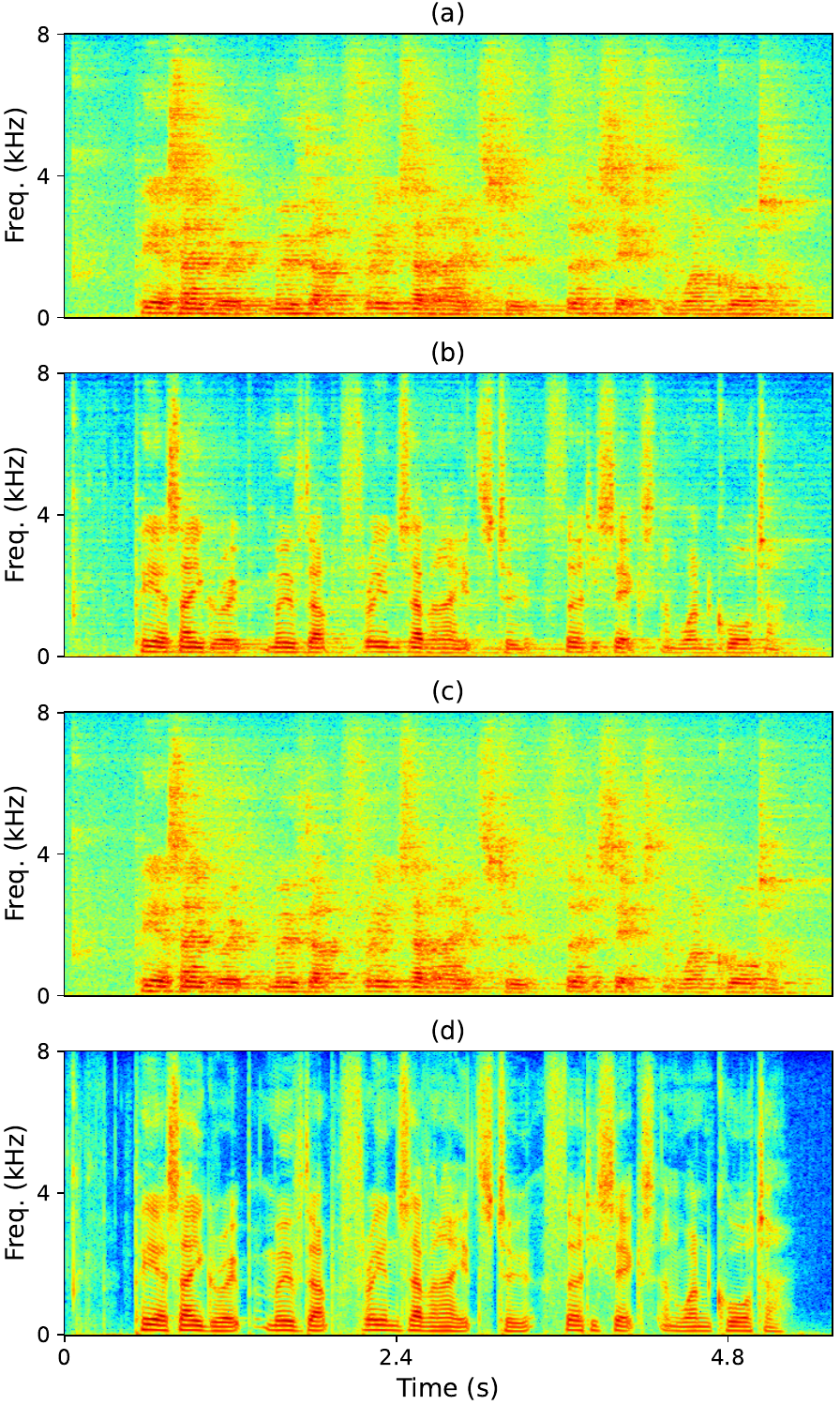}
  \vspace{-0.1cm}
  \caption{
  Illustration of (a) unprocessed mixture; (b)  dereverberation results of USDnet (configured in row 3a of Table \ref{results_garbage_sources}); (c)  dereverberation results of WPE (row 4a of Table \ref{results_garbage_sources}); and (d) clean anechoic speech.
  For this example, the T60 is $1.204$ seconds and direct-to-reverberation energy ratio is around $-5.9$ dB, the PESQ scores are respectively $1.44$, $2.41$, $1.51$ and $4.55$, and eSTOI scores are respectively $0.322$, $0.735$, $0.423$ and $1.0$.
  Best in color.
  }
  \label{dereverb_figure}
\end{figure}

\begin{table*}[t]
\scriptsize
\centering
\captionsetup{justification=centering}
\sisetup{table-format=2.2,round-mode=places,round-precision=2,table-number-alignment = center,detect-weight=true,detect-inline-weight=math}
\caption{\textsc{WER (\%) on Real Data of REVERB (1-channel Input)}}
\vspace{-0.1cm}
\label{results_REVERB_1ch}
\setlength{\tabcolsep}{3pt}
\begin{tabular}{
c %
c
l %
c %
c %
S[table-format=1,round-precision=0] %
S[table-format=1.1,round-precision=1] %
S[table-format=1.2,round-precision=2] %
S[table-format=1.3,round-precision=3] %
S[table-format=1.1,round-precision=1] %
S[table-format=2.1,round-precision=1]
S[table-format=2.1,round-precision=1]
S[table-format=2.1,round-precision=1]
S[table-format=2.1,round-precision=1]
S[table-format=2.1,round-precision=1]
S[table-format=2.1,round-precision=1]
}
\toprule
& & & & & & & & & & \multicolumn{3}{c}{WER (\%)$\downarrow$ on val. set} & \multicolumn{3}{c}{WER (\%)$\downarrow$ on test set} \\
\cmidrule(lr{9pt}){11-13} \cmidrule(lr){14-16}
Row & \multicolumn{1}{c}{Systems} & \multicolumn{1}{c}{Cross-reference} & \#CH in input/loss & $I/J/K/\Delta/L$ & $\alpha$ & $\theta$ &{{PESQ$\uparrow$}} & {{eSTOI$\uparrow$}} & {{SI-SDR (dB)$\uparrow$}} & \multicolumn{1}{c}{Near} & \multicolumn{1}{c}{Far} & {Avg.} & \multicolumn{1}{c}{Near} & \multicolumn{1}{c}{Far} & {Avg.} \\

\midrule

1 & Mixture & - & 1 / - & {-} & {-} & {-} & 1.6406105433337752 & 0.493757274092272 & -3.563633992835417 & 15.35 & 16.88 & 16.11 & 17.09 & 17.29 & 17.19 \\

\midrule

2a & USDnet & 2a of Table \ref{results_simulated_number_of_channels} & 1 / 1 & 40 / 0 / 40 / 3 / - & 1 & {-} & 1.759163199430006 & 0.561225480546709 & -2.1473806877989357 & 21.09 & 24.20 & 22.64 & 20.73 & 22.08 & 21.41 \\

2b & USDnet & 2d of Table \ref{results_simulated_number_of_channels} & 1 / 8 & 40 / 0 / 40 / 3 / - & 1 & {-} & 2.350717771170258 & 0.7420943201511349 & 2.2303781667320615 & 11.73 & 15.04 & 13.38 & 11.56 & 12.63 & 12.10 \\

2c & USDnet & 2 \,\,\,of Table \ref{results_microphone_weight} & 1 / 8 & 40 / 0 / 40 / 3 / - & {$\frac{3}{8-1}$} & {-} & 2.4806705378155853 & 0.7485352414798231 & 2.2246829971965285 & \bfseries 10.29 & \bfseries 12.44 & \bfseries 11.36 & 10.60 & 11.75 & 11.18 \\

2d & USDnet & 2a of Table \ref{results_garbage_sources} & 1 / 8 & 60 / 0 / 60 / 3 / - & {$\frac{3}{8-1}$} & {-} & 2.4463804130682836 & 0.7569215634436278 & 2.495406880906926 & \bfseries 10.29 & 13.33 & 11.81 & 10.19 & 11.51 & 10.85 \\

2e & USDnet & 3a of Table \ref{results_garbage_sources} & 1 / 8 & 60 / 0 / 60 / 3 / 1 & {$\frac{3}{8-1}$} & {-} & \bfseries 2.5270067927005613 & \bfseries 0.7716912101362126 & \bfseries 2.8603948775779355 & 10.36 & \bfseries 12.44 & \bfseries 11.40 & \bfseries 10.06 & \bfseries 10.70 & \bfseries 10.38 \\

\midrule

3a & WPE \cite{Nakatani2010} & 4a of Table \ref{results_garbage_sources} & 1 / - & {-} & {-} & {-} & 1.777453876250657 & 0.5792489171414862 & -1.6990565286957235 & 15.60 & 15.99 & 15.79 & 13.19 & 15.12 & 14.15 \\
3b & DNN-WPE \cite{Kinoshita2017} & 5a of Table \ref{results_garbage_sources} & 1 / - & {-} & {-} & {-} & 2.0220796733016546 & 0.6899447169423142 & 1.9984655773076692 & 14.29 & 16.13 & 15.21 & 12.78 & 14.65 & 13.71 \\

\bottomrule
\end{tabular}%

\vspace{0.3cm}

\scriptsize
\centering
\captionsetup{justification=centering}
\sisetup{table-format=2.2,round-mode=places,round-precision=2,table-number-alignment = center,detect-weight=true,detect-inline-weight=math}
\caption{\textsc{WER (\%) on Real Data of REVERB (8-Channel Input)}}
\vspace{-0.1cm}
\label{results_REVERB_8ch}
\setlength{\tabcolsep}{3pt}
\begin{tabular}{
c %
c
l %
c %
c %
S[table-format=1,round-precision=0] %
S[table-format=1.1,round-precision=1] %
S[table-format=1.2,round-precision=2] %
S[table-format=1.3,round-precision=3] %
S[table-format=1.1,round-precision=1] %
S[table-format=2.1,round-precision=1]
S[table-format=2.1,round-precision=1]
S[table-format=2.1,round-precision=1]
S[table-format=2.1,round-precision=1]
S[table-format=2.1,round-precision=1]
S[table-format=2.1,round-precision=1]
}
\toprule
& & & & & & & & & & \multicolumn{3}{c}{WER (\%)$\downarrow$ on val. set} & \multicolumn{3}{c}{WER (\%)$\downarrow$ on test set} \\
\cmidrule(lr{9pt}){11-13} \cmidrule(lr){14-16}
Row & \multicolumn{1}{c}{Systems} & \multicolumn{1}{c}{Cross-reference} & \#CH in input/loss & $I/J/K/\Delta/L$ & $\alpha$ & $\theta$ &{{PESQ$\uparrow$}} & {{eSTOI$\uparrow$}} & {{SI-SDR (dB)$\uparrow$}} & \multicolumn{1}{c}{Near} & \multicolumn{1}{c}{Far} & {Avg.} & \multicolumn{1}{c}{Near} & \multicolumn{1}{c}{Far} & {Avg.} \\

\midrule

1 & Mixture & - & 8 / - & {-} & {-} & {-} & 1.6406105433337752 & 0.493757274092272 & -3.563633992835417 & 15.35 & 16.88 & 16.11 & 17.09 & 17.29 & 17.19 \\

\midrule

2a & USDnet & 3d of Table \ref{results_simulated_number_of_channels} & 8 / 8 & 40 / 0 / 40 / 3 / - & 1 & 0 & 2.330097478675667 & 0.7296095164191431 & 2.0467920889483717 & \bfseries 8.98 & \bfseries 11.07 & \bfseries 10.03 & 8.88 & 9.52 & 9.20 \\

2b &USDnet & 4 \,\,\,of Table \ref{results_microphone_weight} & 8 / 8 & 40 / 0 / 40 / 3 / - & {$\frac{3}{8-1}$} & 0 & 2.398516595802155 & 0.7449368389728661 & 2.2450837646433173 & 9.17 & 11.89 & 10.53 & \bfseries 8.46 & \bfseries 8.78 & \bfseries 8.62 \\

2c & USDnet & 2c of Table \ref{results_input_microphone_dropout} & 8 / 8 & 40 / 0 / 40 / 3 / - & {$\frac{3}{8-1}$} & 0.7 & 2.4231708638511646 & 0.7462786091427627 & 2.1455779371709514 & 10.36 & 12.03 & 11.20 & 10.28 & 9.32 & 9.80 \\

2d & USDnet & 2b of Table \ref{results_garbage_sources} & 8 / 8 & 60 / 0 / 60 / 3 / - & {$\frac{3}{8-1}$} & 0.7 & 2.4217523200098245 & 0.7595283343159086 & 2.5781748396445607 & 9.86 & 11.96 & 10.91 & 9.65 & 9.79 & 9.72 \\

2e & USDnet & 3b of Table \ref{results_garbage_sources} & 8 / 8 & 60 / 0 / 60 / 3 / 1 & {$\frac{3}{8-1}$} & 0.7 & \bfseries 2.451282974111018 & \bfseries 0.7605576527097211 & \bfseries 2.675615118266085 & 10.67 & 11.35 & 11.01 & 9.45 & 9.62 & 9.54 \\ %

\midrule

3a & WPE \cite{Nakatani2010} & 4b of Table \ref{results_garbage_sources} & 8 / - & {-} & {-} & {-} & 2.0220796733016546 & 0.6899447169423142 & 1.9984655773076692 & 15.47 & 19.34 & 17.41 & 11.82 & 13.94 & 12.88 \\
3b & DNN-WPE \cite{Kinoshita2017} & 5b of Table \ref{results_garbage_sources} & 8 / - & {-} & {-} & {-} & 2.071846921410318 & 0.7240008832744722 & 2.7831842936820994 & 14.66 & 18.66 & 16.66 & 11.72 & 12.83 & 12.28 \\

\bottomrule
\end{tabular}%
\end{table*}

\subsection{Illustration of Dereverberation Results}

Fig. \ref{dereverb_figure} illustrates and compares the dereverberation results of USDnet and WPE, based on a simulated reverberant speech signal sampled from WSJ0CAM-DEREVERB.
We observe that USDnet produces clear suppression of reverberation and clear reconstruction of target speech patterns, while WPE only reduces reverberation slightly and the speech patterns are still very smeared.
This comparison suggests that USDnet can better reduce reverberation than WPE.
A sound demo is provided (see the end of Section \ref{intro}).

\subsection{ASR Results on REVERB}

In Table \ref{results_REVERB_1ch} and \ref{results_REVERB_8ch}, we evaluate USDnet trained on WSJ0CAM-DEREVERB directly on the ASR tasks of REVERB.
Although WSJ0CAM-DEREVERB is simulated and is mismatched with the real-recorded REVERB dataset, we observe that USDnet can significantly improve ASR performance in both near- and far-field cases over unprocessed mixtures.
In both tables, USDnet outperforms WPE and DNN-WPE.

In Table \ref{results_REVERB_8ch}, we observe that applying input microphone dropout (see the $\theta$ column), although producing better enhancement scores, is detrimental to ASR.

Comparing the results in Table \ref{results_REVERB_1ch} and \ref{results_REVERB_8ch}, we observe that USDnet with eight-channel input obtains better recognition performance but worse PESQ, eSTOI and SI-SDR scores than the one with monaural input.

\ZQHL{
\section{Miscellaneous Discussions}\label{Miscellaneous_Discussions}

This section discusses USDnet in aspects other than dereverberation performance, putting USDnet is a broader context.

\subsubsection{Algorithmic Complexity}

As described in Section \ref{run_time_inference}, at run time the trained model performs inference in the same way as existing supervised dereverberation algorithms (i.e., just running a forward pass to obtain target estimate $\hat{S}_q$).
During training, the extra amount of computation compared to supervised algorithms stems from estimating FCP filters and computing the $\mathcal{L}_{\text{MC}}$ loss, which are typically much less costly than modern DNN modules.

\subsubsection{Amenability to Online Dereverberation}

USDnet can be easily configured for online derevebreration. We only need to change the DNN to be causal.
We leave this investigation to future research.

\subsubsection{Array-geometry-invariant Modeling}\label{invariant_Description}

USDnet can be modified to be trained on mixtures recorded by arrays with diverse microphone geometries.
Clearly, the $\mathcal{L}_{\text{MC}}$ loss is invariant to microphone geometry.
We only need to use a DNN architecture that can model mixtures with a diverse number of input microphones, by following, e.g., the VarArray approach \cite{Yoshioka2022VarArray}.
We leave this to future research.
}

\ZQHL{
\section{Limitations}\label{limitation}

A limitation of USDnet stems from the assumption that there is only one target speaker source and the considered noises are weak and stationary.
In realistic conditions (such as in conversational applications), there could be an unknown number of target speaker sources and an unknown number of directional and diffuse sources, and we need to address simultaneous separation, dereverberation and enhancement.
Future research will modify USDnet for these more challenging cases.
Our recent studies such as SuperM2M \cite{Wang2024SuperM2M} and cross-talk reduction \cite{Wang2023CTRnet} have made initial investigation along this direction.

Another limitation is that, in our experiments, multi-channel USDnet trained with, e.g., eight-channel input and loss shows worse performance than single-channel USDnet trained with, e.g, single-channel input and eight-channel loss.
This issue is likely because in the case of eight-channel input and loss, USDnet sees the same mixtures used in loss computation from the input, and hence could figure out a way to optimize the loss well but not dereverberate well.
This issue needs to be addressed in future research to build better multi-channel models.
}

\section{Conclusion}\label{conclusion}

We have formulated unsupervised neural speech dereverberation as a blind deconvolution problem, which requires estimating both the source and filter, and proposed a novel algorithm named USDnet, which leverages over-determined training mixtures as constraints to narrow down the solutions to target speech.
Evaluation results on two dereverberation tasks show that USDnet, even if only trained on a set of reverberant mixtures in an unsupervised way, can learn to effectively reduce reverberation.
Future research will modify USDnet and evaluate it on real-recorded conversational speech dereverberation and separation tasks with multiple sound sources.

A contribution of this paper, we highlight, is our finding that DNNs can be trained directly on a set of reverberant mixtures, in an unsupervised way, to reduce reverberation, by solely exploiting mixture constraints afforded by multi-microphone mixtures.
The novel way of modeling reverberation and the novel concept of leveraging mixture constraints exploit signal cues and physical principles very different from the ones utilized by existing algorithms.
The proposed algorithms, hence, we think, would in many ways complement existing dereverberation and separation algorithms, and motivate new ones.

In closing, another contribution of this paper, we point out, is our unsupervised deep learning based approach for solving blind deconvolution problems, which exist not only in speech dereverberation but also in many other engineering applications.
By modeling source priors in an unsupervised way, filter estimation becomes a differentiable operation given a source estimate.
This enables us to train DNNs on signal patterns in a discriminative way to optimize mixture-constraint loss and realize unsupervised deconvolution.
This novel methodological contribution, we think, could generate broader impact beyond speech dereverberation.

\bibliographystyle{IEEEtran}
\bibliography{references.bib}

\begin{IEEEbiography}[{\includegraphics[width=1in,height=1.25in,clip,keepaspectratio]{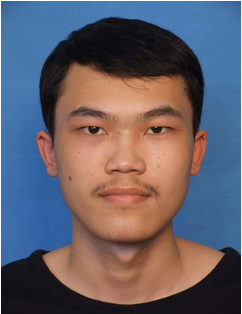}}]{Zhong-Qiu Wang}
received the B.E. degree in computer science and technology from Harbin Institute of Technology, Harbin, China, in 2013, and the M.S. and Ph.D. degrees in computer science and engineering from The Ohio State University, Columbus, OH, USA, in 2017 and 2020, respectively.
He is currently a tenure-track associate professor in the Department of Computer Science and Engineering at Southern University of Science and Technology, Shenzhen, Guangdong, China.
He was a Postdoctoral Research Associate with Carnegie Mellon University, Pittsburgh, PA, USA, from $2021$ to $2024$, and a visiting research scientist at Mitsubishi Electric Research Laboratories, Cambridge, MA, USA, from $2020$ to $2021$.
His research interests include speech separation, robust automatic speech recognition, microphone array processing, and deep learning, aiming at solving the cocktail party problem.
\end{IEEEbiography}

\end{document}